\documentclass[12pt]{article}

\input{epsf}
\usepackage[nosort]{cite}
\usepackage{cite}
\usepackage{amsmath,pifont,bbding}
\usepackage{epsfig}
\usepackage{amsfonts}
\usepackage{amssymb}
\usepackage{multirow}
\usepackage{enumerate}
\usepackage{subfigure}
\usepackage{slashed}
\usepackage{color}

\usepackage{subfigure}

\usepackage{caption}

\newcommand{\be}{\begin{equation}}
\newcommand{\ee}{\end{equation}}
\newcommand{\bee}{\begin{equation*}}
\newcommand{\eee}{\end{equation*}}
\newcommand{\bea}{\begin{eqnarray}}
\newcommand{\eea}{\end{eqnarray}}
\newcommand{\bean}{\begin{eqnarray*}}
\newcommand{\eean}{\end{eqnarray*}}

\begin{document}

\setcounter{page}{1}
\thispagestyle{empty}

\begin{flushright}
October 16, 2018 \\
DESY 18-109
\end{flushright}

\vskip 8pt

\begin{center}
{\bf \Large{High Scale Electroweak Phase Transition:   \\
\vskip 12pt
Baryogenesis \& Symmetry Non-Restoration
}}
\end{center}

\vskip 12pt

\begin{center}
 {\bf  Iason Baldes$^{a}$ and G\'eraldine Servant$^{a,b}$}
 \end{center}

\vskip 14pt

\begin{center}
\centerline{$^{a}${\it DESY, Notkestra{\ss}e 85, D-22607 Hamburg, Germany}}
\centerline{$^{b}${\it II. Institute of Theoretical Physics, University of Hamburg, D-22761 Hamburg, Germany}}

\vskip .3cm
\centerline{\tt iason.baldes@desy.de, geraldine.servant@desy.de}
\end{center}

\vskip 10pt

\begin{abstract}
\vskip 3pt
\noindent

We explore the possibility that the electroweak phase transition happens at a scale
much higher than the electroweak scale today.
In this context, high scale CP-violating sources for electroweak baryogenesis 
are not constrained by low-energy experiments.
We propose a scenario of high-scale electroweak baryogenesis linked to flavour physics. This scenario allows for a period of
enhanced Yukawa couplings during the evolution of the universe, which source time-dependent CP violation.
The electroweak symmetry is never restored after the high-scale phase transition due to negative contributions to the Higgs thermal mass squared from a large number of additional electroweak-scale neutral scalars coupling to the Higgs. As a result,
the washout of the baryon asymmetry is avoided.

\end{abstract}

\newpage

\tableofcontents

\vskip 13pt


\section{Introduction}
Two leading theories for explaining the matter-antimatter asymmetry are leptogenesis~\cite{Fukugita:1986hr} and electroweak baryogenesis (EWBG)~\cite{Shaposhnikov:1987tw,Cohen:1990it}. The former is intrinsically tied to the mass scale of the Majorana neutrinos, $M_{N}$, typically taken to be at a very high scale.
In standard, non-flavoured, thermal leptogenesis with a hierarchy between the lightest and heavier sterile states, the Davidson-Ibarra bound requires $M_{N} \gtrsim 10^{9}$ GeV~\cite{Davidson:2002qv}.
Nevertheless, detailed work has shown that the mass scale can be reduced, even down to $M_{N} \sim 1$ GeV, when flavour effects and the possibility of resonant enhancements with quasi-degenerate spectra are taken into account~\cite{Akhmedov:1998qx}.

In contrast, it is usually assumed EWBG is intrinsically tied to the scale of electroweak symmetry breaking $\sim 100$ GeV. This is positive from the viewpoint of testability, as the required CP violation can be constrained from measurements of electric dipole moments (EDMs)~\cite{Baron:2013eja}, and the requirement for a strong first order phase transition typically leads to sufficient modifications of the zero-temperature Higgs potential for deviations to be measurable at colliders in the near future~\cite{Grojean:2004xa,Curtin:2014jma,DiVita:2017eyz,DiVita:2017vrr}. This has led to a healthy tightening of the constraints on the scenario in the past few decades. Recently, the idea of linking EWBG to the flavour sector has been advocated. This has the advantage of possibly: (i) providing the CP violation required for EWBG while making the CP violation time dependent~\cite{Berkooz:2004kx,Perez:2005yx,Hall:2005aq,Baldes:2016rqn,Baldes:2016gaf,vonHarling:2016vhf,Bruggisser:2017lhc,Bruggisser:2018mus,Bruggisser:2018mrt,Servant:2018xcs} and hence naturally evading the EDM bounds, (ii) making the phase transition strong through the varying Yukawa couplings~\cite{Baldes:2016rqn,Baldes:2016gaf}. Bringing flavour constraints into the game, however, makes model building in this framework challenging. It would therefore be helpful to raise the scale of EWBG, so we can in turn also raise the flavour scale and hence more easily satisfy the flavour constraints. 

\begin{figure}[t!]
\begin{center}
\includegraphics[height=200pt]{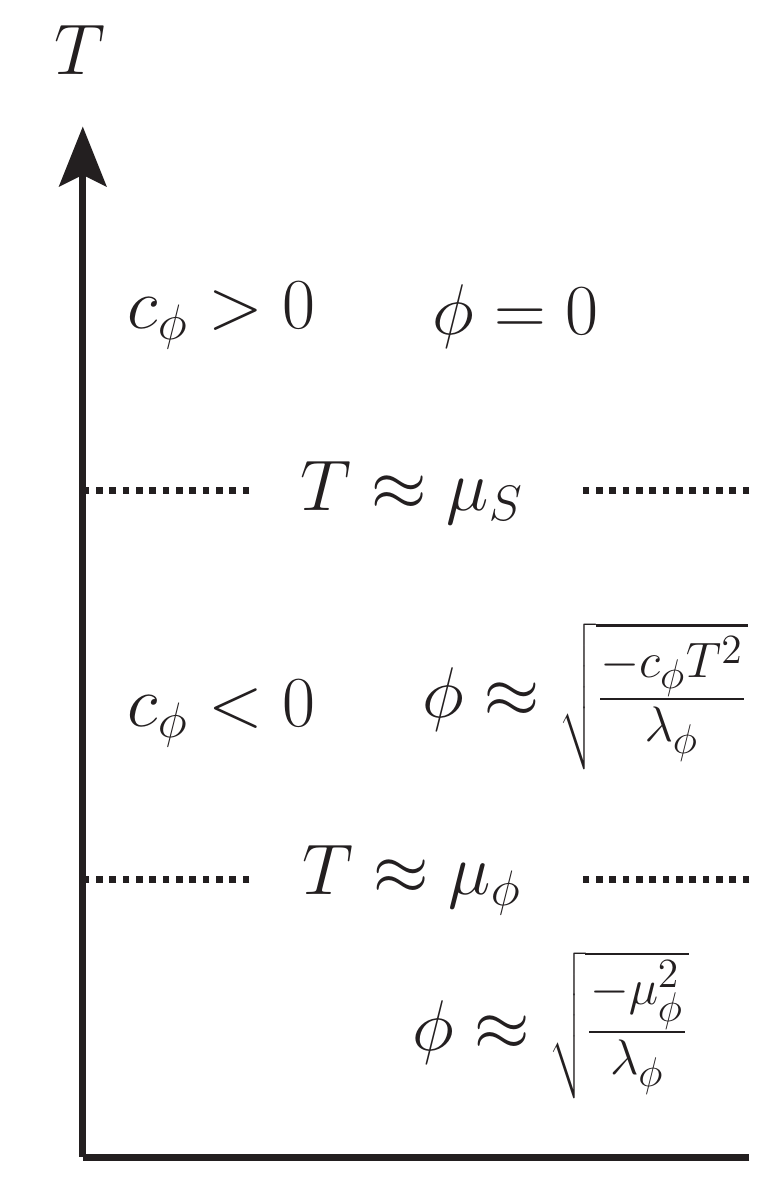} \\
\end{center}
\caption{\small Sketch of the effect illustrated in the toy model. At high temperature the thermal mass of $\phi$, $c_{\phi}T^2$, is positive and the VEV is zero. The temperature drops below a mass threshold of a field $S$, removing a positive contribution to the thermal mass of $\phi$. The thermal mass of $\phi$ is then negative due to the contributions from some additional scalars $\chi_{i}$ and the VEV becomes proportional to the temperature. Finally, at sufficiently low temperatures, the VEV is set by the usual minimization conditions of the zero temperature potential.}
\label{fig:toymodelsketch}
\pagestyle{empty}
\end{figure}

More broadly, raising the scale of EW symmetry breaking is anyway an exciting theoretical possibility, not limited to the context of the flavour model considered below. The aim of this paper therefore is to study the possibility of high scale EWBG, in which the Higgs $\phi$  first obtains a large vacuum expectation value (VEV), which is later gradually decreased to $v_{\phi} = 246$ GeV while in the broken electroweak phase. The VEV can be gradually decreased using a symmetry non-restoration effect, in which the Higgs --- through the coupling to other scalar fields --- gains a negative thermal mass squared and hence a VEV proportional to the temperature~\cite{Weinberg:1974hy,Mohapatra:1979qt,Fujimoto:1984hr,Salomonson:1984rh,Salomonson:1984px,Dvali:1995cj,Bimonte:1995xs,Bimonte:1995sc,Dvali:1996zr,Orloff:1996yn,Pietroni:1996zj,Gavela:1998ux}.\footnote{For brevity, we omit ``squared" when discussing the thermal masses of scalar fields from now on.} In the models of symmetry non-restoration considered so far, the broken symmetry is not restored at any temperature. For electroweak baryogenesis, however, we want the Higgs to start in the symmetric phase and undergo a phase transition into the broken phase. Here, we will first show the two conditions can be realised together generically, through a simple toy model example, sketched in Fig.~\ref{fig:toymodelsketch}. 

Motivated by our findings, we then return to flavour considerations in a more complete model, in which the Yukawa couplings are field-dependent and large at early times. The flavor sector contains extra fermions whose mass is controlled by the VEV of a scalar field $\Delta$ that sets the flavour scale, $\gtrsim {\cal O}$(10) TeV, today. The broken EW phase minimum develops at large Higgs values once the temperature drops to the flavour scale. The Higgs then undergoes a strong first order phase transition from a point in field space in which the Yukawa couplings are $\mathcal{O}$(1) --- allowing for enhanced CP violation compared to the SM --- into the broken phase minimum~\cite{Berkooz:2004kx,Perez:2005yx,Hall:2005aq,Baldes:2016rqn,Baldes:2016gaf,vonHarling:2016vhf,Bruggisser:2017lhc,Bruggisser:2018mus,Bruggisser:2018mrt}. 
This is when baryogenesis takes place.

\begin{figure}[t!]
\begin{center}
\includegraphics[width=200pt]{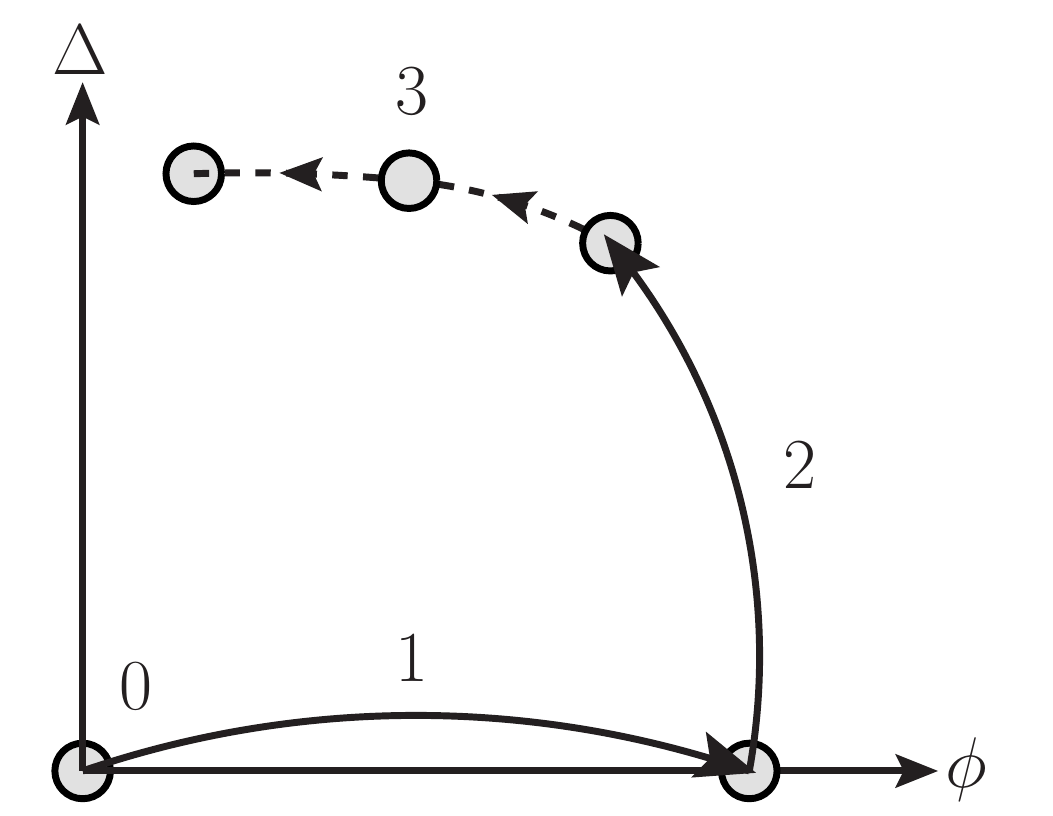} \\
\end{center}
\caption{\small Summary of the pattern of the phase transitions in the full model. Step 0: the fields start in the symmetric phase. Step 1: the strong first order EW phase transition. Step 2: the first order phase transition in the $\Delta$ direction during which the Yukawa couplings approach their present values. Step 3: evolution of the minimum with temperature in which $\phi$ decreases due to the symmetry non-restoration effect and $\Delta$ increases due to falling thermal support.}
\label{fig:potevodiagram}
\pagestyle{empty}
\end{figure}

Through another phase transition the Yukawa couplings are suppressed to their present day values. The Higgs also obtains a negative thermal mass and the VEV of the Higgs gradually decreases to $v_{\phi} = 246$ GeV as the temperature drops. The washout avoidance condition, $\phi/T \gtrsim 1$, is maintained throughout the evolution of the potential following the first phase transition. The sequence of phase transitions is summarised in Fig.~\ref{fig:potevodiagram}. The scenario is a novel realisation of high scale electroweak baryogenesis linked to flavour physics.

The working model we present here is intended as an initial exploration of such ideas, through which model building problems can be identified and hopefully serve as inspiration for future work in more realistic contexts. The scenario shares some aspects of the leptogenesis scenario in Ref.~\cite{Hamada:2016gux}. Though in that model --- as the symmetry non-restoration effect is not used --- EW symmetry is restored in an intermediate step. Hence, as was noted by the authors, it would require an additional source of $B-L$ number violation if it were to be modified for EWBG. 

The paper is organised as follows. In the next section we illustrate the idea of a high scale phase transition or crossover using the symmetry non-restoration effect in a toy model. In Sec.~\ref{sec:SectionModel} we describe the full model, starting from the fermionic sector and flavour structure and then moving onto the scalar potential and the symmetry non-restoration effect. In Sec.~\ref{sec:phasetrans} we discuss the details of the phase transition. 
In Sec.~\ref{sec:relaxation}, we briefly comment on a possible earlier cosmological history to justify our choice of potential parameters. 
We then discuss related phenomenology in Sec.~\ref{sec:pheno}, namely possible gravitational signals and the constraints on the low mass scalar states required by the symmetry non-restoration effect, before concluding our discussion.

\section{Toy Example}
\label{sec:ToyModel}

High temperature symmetry non-restoration was studied some time ago~\cite{Weinberg:1974hy,Mohapatra:1979qt,Fujimoto:1984hr,Salomonson:1984rh,Salomonson:1984px,Dvali:1995cj,Bimonte:1995xs,Bimonte:1995sc,Dvali:1996zr,Orloff:1996yn,Pietroni:1996zj,Gavela:1998ux}, mainly in the context of GUT theories or in the context of SUSY flat directions~\cite{Bajc:1998jr}.
The phenomenon has been confirmed by lattice simulations~\cite{Jansen:1998rj,Bimonte:1999tw} and non-perturbative methods~\cite{Pinto:1999pg}.
For  the electroweak symmetry, it was considered only a few times. The possible existence of a broken phase of electroweak symmetry at high temperature in Little Higgs extensions of the Standard Model was investigated in~\cite{Espinosa:2004pn,Aziz:2009hk}. The theory, however, exhibits a restoration of electroweak symmetry as long as temperatures are not pushed beyond the range of validity of the EFT for a finite temperature calculation \cite{Ahriche:2010kh}. This conclusion is generalised to Twin Higgs models in \cite{Kilic:2015joa} and confirms earlier findings in \cite{Gavela:1998ux}.
The case of composite Higgs models with partial fermion compositeness in which the Higgs is a PNGB has been studied recently  in and these models also lead to EW symmetry restoration~\cite{Bruggisser:2018mus,Bruggisser:2018mrt}.

Here we will implement the ideas illustrated in Fig.~\ref{fig:toymodelsketch}, and show how a phase transition or crossover can occur at a high scale, i.e.~above the zero-temperature minimum of the scalar potential, using an extension of the symmetry non-restoration effect. Unlike in earlier realisations of the symmetry non-restoration effect, the symmetry is actually restored at a sufficiently large temperature, i.e. above some mass threshold. Here, by symmetry non-restoration, we mean that at temperatures below the phase transition one of the scalar fields obtains a VEV proportional to the temperature. 

The main idea is to induce a negative thermal mass for the Higgs through a negative cross-quartic coupling between the Higgs and a large number of additional scalar fields.
Consider a toy model of scalar fields, $\phi$, $S$, and $\chi_{i}$, where $i=1,...,N_{\rm Gen}$ is a generational index (the reason for considering multiple generations will be made clear below). We denote the degrees of freedom with $N_{\phi}$, $N_{S}$, and $N_{\chi_i}$ (the $\chi$ sector therefore has in total $N_{\chi} = N_{\rm Gen}N_{\chi_i}$ degrees of freedom). In this section $\phi$ is acting as a placeholder for the EW Higgs, though we switch off the usual SM Yukawa and gauge interactions for the discussion in this section. For the purposes of our example, the relevant terms in the tree level potential are given by
  	\begin{equation}
	V(\phi,\chi) = \frac{\mu_{S}^{2}}{2}S^{2} + \frac{\mu_{\chi}^{2}}{2}\sum_{i}\chi_{i}^{2} + \frac{\mu_{\phi}^{2}}{2}\phi^{2} + \frac{\lambda_{\phi}}{4}\phi^{4} + \frac{\lambda_{\chi}}{4}\sum_{i}\chi_{i}^{4} + \frac{\lambda_{S}}{4} S^{4} + \frac{\lambda_{\phi \chi}}{4}\phi^{2}\sum_{i}\chi_{i}^{2} + \frac{\lambda_{\phi S}}{4}\phi^{2}S^{2},
	\label{eq:toypotential}
	\end{equation}
where for simplicity we assume degenerate masses and couplings for the $\chi_{i}$ generations and that the cross quartic $\lambda_{\chi S}$ is negligible. As we shall be choosing $\lambda_{\phi \chi} <0$, stability of the tree level potential requires
	\begin{equation}
	\lambda_{\phi \chi} > -2 \sqrt{ \frac{ \lambda_{\phi}\lambda_{\chi} }{ N_{\rm Gen} } }.
	\end{equation}
At high temperatures, $T \gg \mu_{\phi}, \; \mu_{\chi}$, the thermal masses of the fields are~\cite{Katz:2014bha}
	\begin{align}
	c_{\chi_{i}}T^{2} & \approx \left( \left[N_{\chi_{i}}+2\right]\frac{\lambda_\chi}{12} + N_{\phi}\frac{ \lambda_{\phi\chi} }{ 24 } \right)T^{2}, \label{eq:chithermalmass} \\
	c_{S}T^{2} & \approx \left( \left[N_{S}+2\right]\frac{ \lambda_S }{ 12 } + N_{\phi}\frac{ \lambda_{\phi S} }{ 24 } \right)T^{2}, \\
	c_{\phi}T^{2} & \approx
		 \begin{cases}
	 \left( \left[N_{\phi}+2\right]\frac{\lambda_\phi}{12} + N_{\chi}\frac{ \lambda_{\phi\chi} }{ 24 } + N_{S}\frac{ \lambda_{\phi S} }{ 24 } \right)T^{2} & \text{for} \; T \gtrsim \mu_{S}, \\
	\left( \left[N_{\phi}+2\right]\frac{\lambda_\phi}{12} + N_{\chi}\frac{ \lambda_{\phi\chi} }{ 24 }  \right)T^{2} & \text{for} \; T \lesssim \mu_{S}.
		 \end{cases}
	\label{phithermalmasstoyexample}
	\end{align}
Now consider a judicious choice of parameters so that: (i) $\chi_{i}$ and $S$ always have positive thermal masses, (ii) $c_{\phi}$ is positive at high temperature, (iii) $c_{\phi}$ becomes negative when the contribution of $S$ to its thermal mass becomes negligible, i.e. once $T \lesssim \mu_{S}$. The effective potential in the $\phi$ direction, when $T \gg \mu_{\phi}$ can be approximated as $c_{\phi}T^{2} \phi^{2}/2+ \lambda_{\phi}\phi^{4}/4$. Positive $c_{\phi}$ returns a minimum at $\phi=0$, but for negative $c_{\phi}$ we will find a minimum at $\phi = \sqrt{c_{\phi}/\lambda_{\phi}}T$. The latter solution is the usual symmetry non-restoration effect~\cite{Weinberg:1974hy,Mohapatra:1979qt,Fujimoto:1984hr,Salomonson:1984rh,Salomonson:1984px,Dvali:1995cj,Bimonte:1995xs,Bimonte:1995sc,Dvali:1996zr,Gavela:1998ux}. What is new here is the presence of the additional field $S$ which can switch the sign of $c_{\phi}$ when $T$ reaches a mass threshold, leading to a phase transition or crossover. (Similarly, the symmetry non-restoration effect disappears if $T$ falls sufficiently below $\mu_{\chi}$.) Eventually, for $T \ll |\mu_{\phi}|$, the VEV is set by the usual zero-temperature minimization conditions.

\begin{figure}[t]
\begin{center}
\includegraphics[width=220pt]{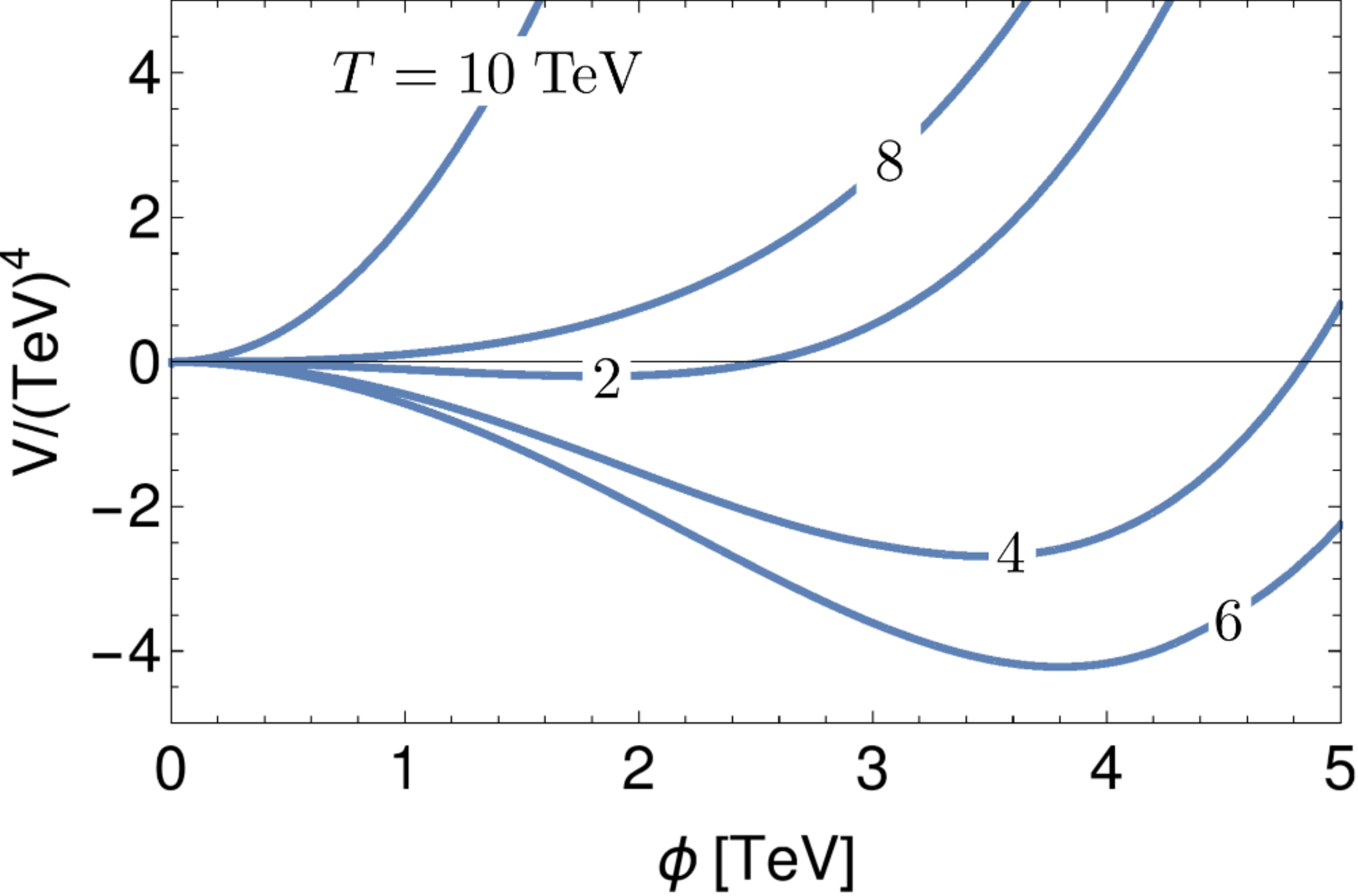}
\includegraphics[width=220pt]{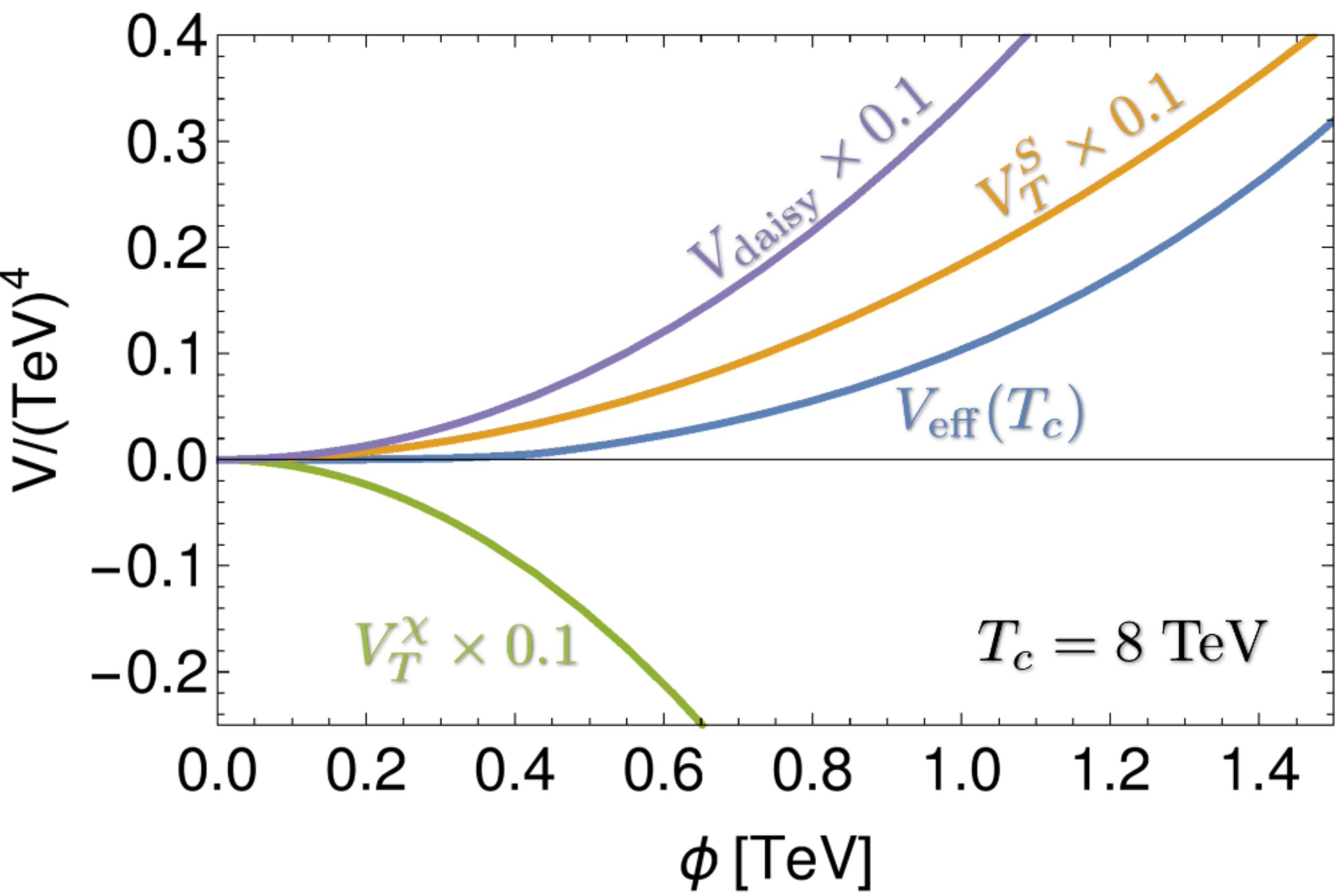}
\end{center}
\caption{\small Left: The evolution of the effective potential with the temperature in the toy model showing a crossover at $T_{c} \approx 8$ TeV. Right: The effective potential in the toy model at $T_{c} \approx 8$ TeV. The positive thermal contributions from the daisy resummation and $S$, and the negative thermal contribution from the $\chi_{i}$ are also shown.}
\label{fig:toyexample}
\end{figure}

\begin{figure}[t]
\begin{center}
\includegraphics[width=220pt]{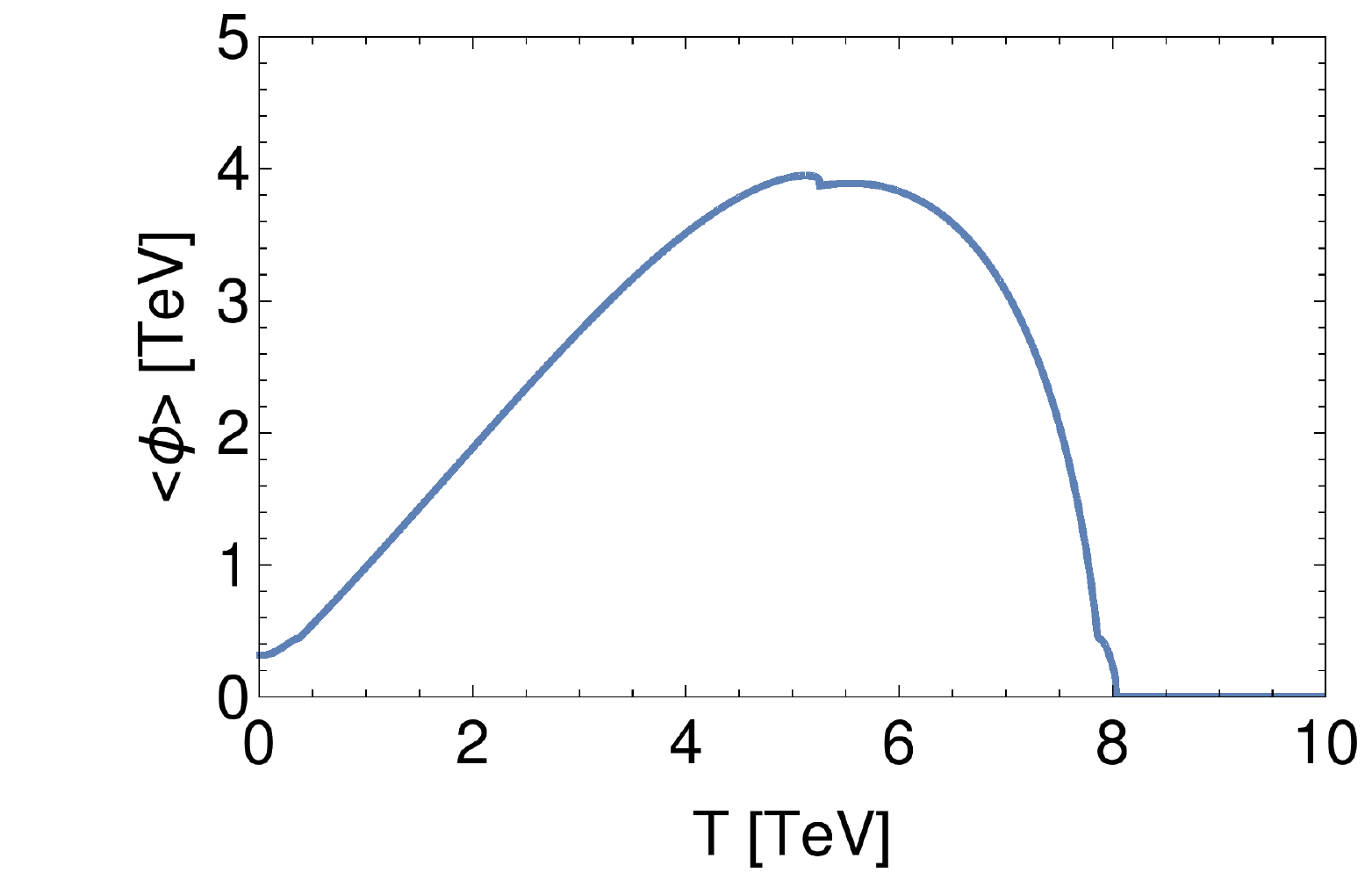}
\includegraphics[width=220pt]{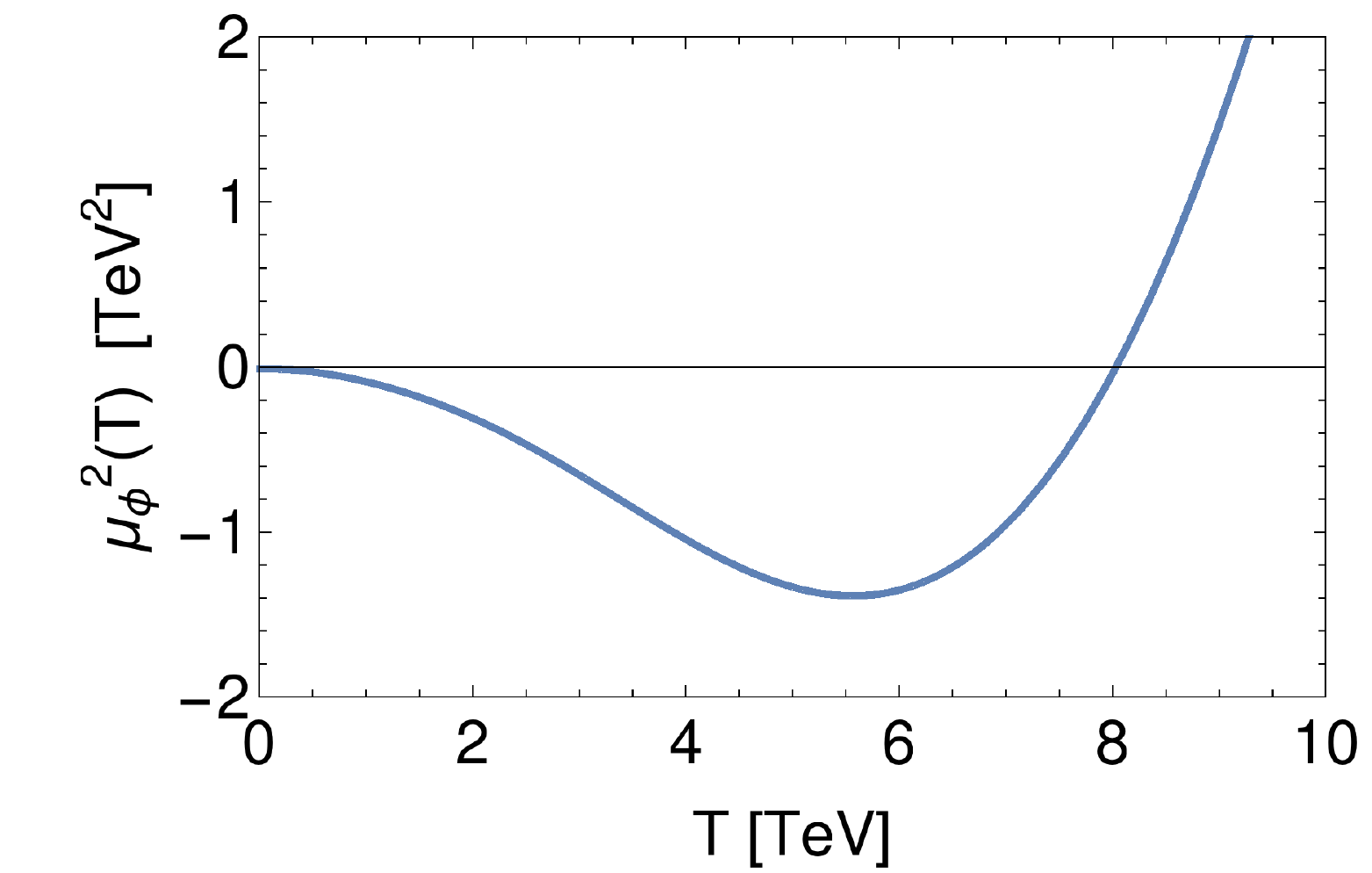}
\end{center}
\caption{\small Left: The VEV of $\phi$ as a function of $T$ in the toy model. Right: The effective mass squared of the $\phi$, i.e. the second derivative of the potential, at the origin in field space.}
\label{fig:toyexample2}
\end{figure}

We numerically evaluate the effective potential including the tree-level terms, zero and finite-temperature one-loop terms, and the daisy resummation.\footnote{We use the Arnold-Espinosa method of implementing the daisy resummation~\cite{Arnold:1992rz}. We cut off the contribution of $S$ to the thermal masses with an exponential factor, $e^{-m_{S}/T}$, in order to avoid spurious contributions to the daisy resummation. We checked that the thermal
mass estimated using the high-temperature expansion is consistent 
with the second derivative of the one loop thermal terms. In fact, the phase transition is stronger when using the numerical value rather than the high-temperature expansion value.} 
The latter is crucial and weakens the phase transition. 
To give a concrete example, consider the choice of parameters\footnote{Motivated by flavour bounds, we take a characteristic scale $\mu_S\sim {\cal O}(10) $ TeV for illustration. The scale of the transition, however, can be taken much larger. The main limit for baryogenesis is around $T\sim 10^{12}$ GeV when the sphalerons become out-of-equilibrium in the symmetric phase.}
	\begin{gather}	
	N_{\phi} = 1, \qquad  N_{\rm Gen}  = 12, \qquad   N_{\chi_i} = 4, \qquad N_{S} = 12,  \nonumber \\
	\lambda_{\phi}  = 0.1, \qquad   \lambda_{\chi}  = 0.5, \qquad  \lambda_{S}=1, \qquad \lambda_{ \phi \chi}  = -0.1, \qquad   \lambda_{ \phi S}  =  1,   \\ 
	 \mu_{\phi}  = i \times 0.1 \; \mathrm{TeV}, \qquad \mu_{\chi} = 0.1 \; \mathrm{TeV},  \qquad \mu_{S} = 20 \; \mathrm{TeV}. \nonumber 
	\end{gather}
In Fig.~\ref{fig:toyexample} we show the resulting cross over, together with the thermal contributions from the $S$ and $\chi_{i}$ scalars and the daisy resummation. In Fig.~\ref{fig:toyexample2} we plot the evolution of the VEV and effective mass of $\phi$ as a function of $T$, showing the various stages discussed above. As mentioned previously, the mass threshold is naively at $T \sim \mu_{S}$, however, additional factors which enter the full expressions lead to the non-zero VEV only developing at $T \approx \mu_{S}/2$ in our example. We have checked the $\chi_{i}$ VEVs remain zero throughout due to positive thermal contributions in the $\chi_{i}$ field directions.

The reason for requiring multiple generations of $\chi_{i}$ is revealed by considering the thermal mass of the $\chi_{i}$, Eq.~(\ref{eq:chithermalmass}). A large thermal mass spoils the symmetry non-restoration effect once it enters the effective potential through the daisy resummation~\cite{Fujimoto:1984hr}. This is because a large thermal mass can make the vacuum contribution, $-\lambda_{\phi \chi}\phi^{2}/2$, which leads to the symmetry non-restoration effect, negligible in the effective potential. (This is not captured in the naive Eq.~(\ref{phithermalmasstoyexample}) which is simply based on a high-$T$ expansion.) Assuming, as we do, that $\lambda_{\chi} > \lambda_{\phi}$, the use of multiple generations means the thermal mass of the $\chi_{i}$ can be reduced, assuming the inter-generational interactions are negligible. Thus allowing for the symmetry non-restoration phase to proceed even once the daisy resummation is included. Because of the different multiplicities and couplings, the two-loop thermal masses are parametrically suppressed compared to the one-loop thermal masses, thus giving credence to our perturbative analysis. Furthermore, the use of multiple generations allows us entertain the possibility that the $\chi_{i}$ are singlet fields, i.e. $N_{\chi_i}=1$, in our full model below, which leads to simpler low energy phenomenology. 

It is interesting that the stability constraint implies
	\begin{equation}
	\left| N_{\chi}\frac{ \lambda_{\phi\chi} }{ 24 } \right| < \frac{ N_{\chi i} \sqrt{ N_{\rm Gen} \lambda_{\phi} \lambda_{\chi} }} { 12 },
	\end{equation}
which reveals that a negative thermal mass can be achieved for a sufficiently large $N_{\rm Gen}$, while keeping $c_{\chi_{i}}$ small enough, and the potential stable.

In our example here, we do not have a first order phase transition required for EWBG. Nevertheless, we shall see below that in our full model a strong enough phase transition can be achieved. What is important here is that we can start in the symmetric phase at high temperature and make a transition to a period in which $\phi$ obtains a large temperature-dependent VEV. We can then use the additional freedom gained, e.g. by introducing additional field directions, to arrange for a strong first order electroweak phase transition at a high scale followed by the use of the symmetry non-restoration effect to avoid washout.

\section{Complete model}
\label{sec:SectionModel}

To realise the sequence of events described in the introduction, we need two main ingredients: First, there should be negative contributions to the Higgs thermal mass to enable the Higgs potential to exhibit a period of symmetry non-restoration. This can be achieved by adding extra scalar fields coupling to the Higgs as we have just seen in the previous section. Second, for a successful implementation of the EWBG mechanism, we need to start in the symmetric phase, at the highest temperatures. As we will see, we realise this through thermal effects from additional fermions together with higher dimensional terms in the potential, rather than scalar degrees-of-freedom as in the toy example. 

In the following, we start discussing the new fermionic degrees of freedom connected to the flavour sector, as motivated in Froggatt-Nielsen models~\cite{Froggatt:1978nt}. These will provide positive thermal contributions to the effective potential and will also be responsible for creating a minimum in the Higgs potential at large Higgs values, through zero temperature one-loop effects (hence related to step 1 and 2 in Fig.~\ref{fig:potevodiagram}).\footnote{Although the fermions create a minimum at large field values, we cannot trap the EW Higgs at this point to temperatures below the EW scale today without: (i) huge fine tuning of the polynomial potential, (ii) diluting the baryon asymmetry due to the false vacuum energy becoming dominant. Hence the symmetry non-restoration effect is still required to avoid washout.} Furthermore, these fermions help us achieve a strong first order phase transition and are also our source of CP violation~\cite{Baldes:2016rqn,Baldes:2016gaf}.

We then discuss the extended scalar sector consisting of the EW higgs together with a scalar $\Delta$ which controls the masses of the exotic fermions. The phase transition in the $\Delta$ direction is essential in reducing the effective Yukawa couplings to their SM values (step 2). Finally we discuss the new scalars leading to EW symmetry non-restoration, which provide a negative thermal contribution for the first phase transition (step 1), and also enable the final stage of the mechanism (step 3). Combining all these effects, we obtain the more complicated sequence of phase tranisitions sketched in Fig.~\ref{fig:potevodiagram}, in contrast to the transition in a single field direction as in the toy model.

\subsection{The fermionic sector}

\begin{figure}[t]
\begin{center}
\includegraphics[width=220pt]{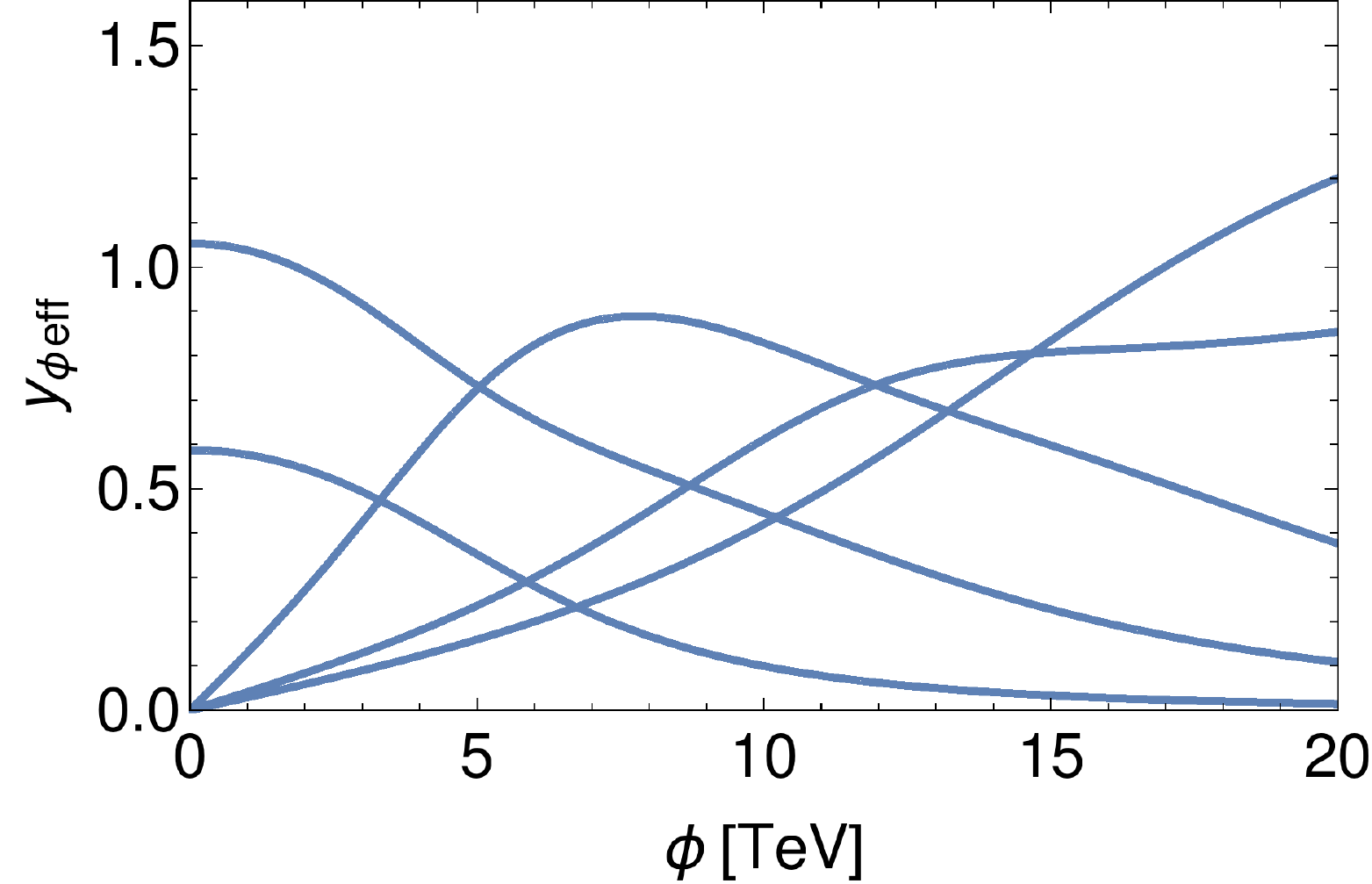} \qquad
\includegraphics[width=220pt]{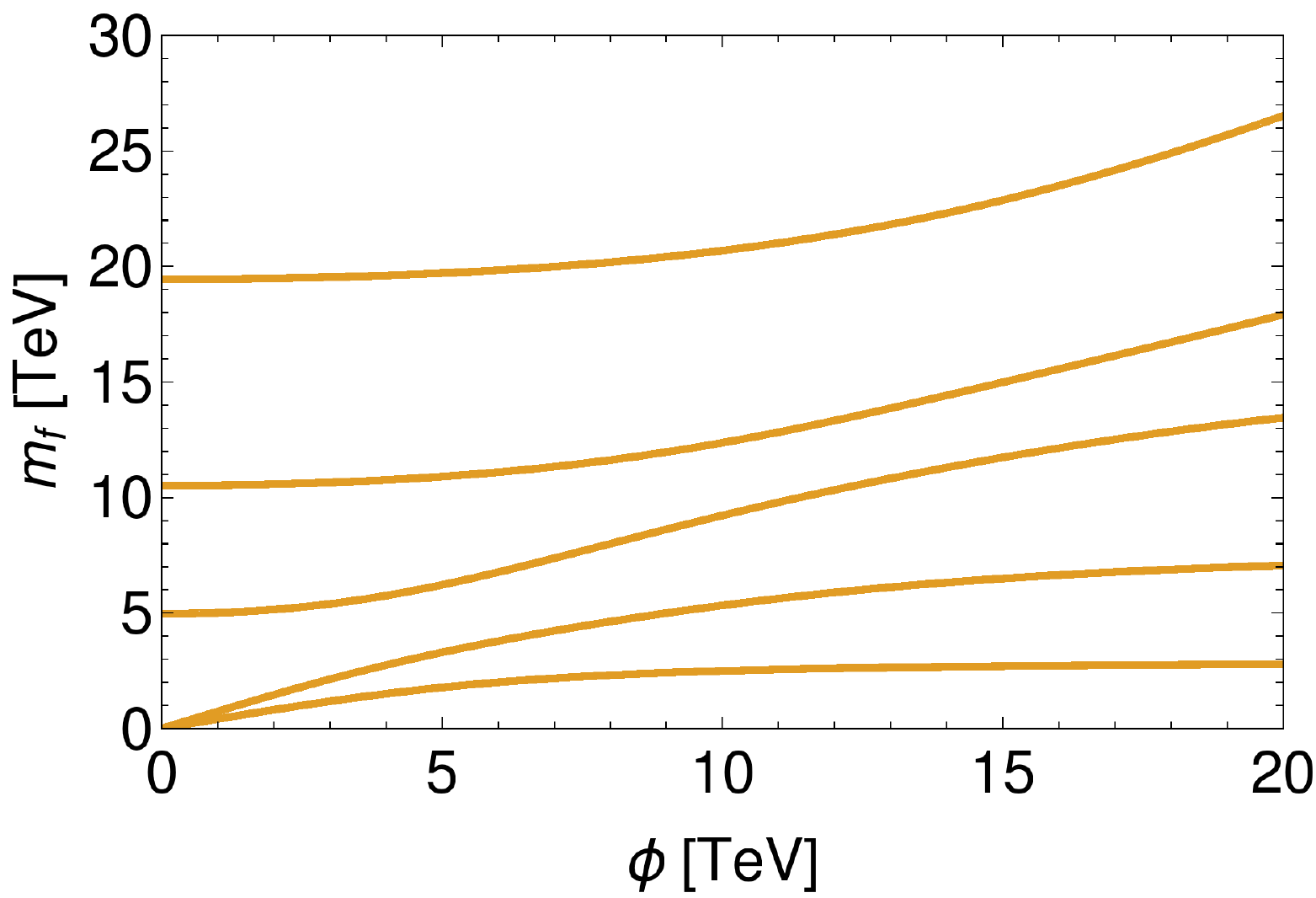}
\end{center}
\caption{\small Left: The effective yukawa couplings of the electroweak Higgs to the fermions along the $\phi$ axis, i.e. $\Delta=0$. Here we have set $a_{s} = 10$ TeV which implies $v_{\Delta} = 50$ TeV. Right: The masses of the fermions along the same path. For the Higgs VEV equal to its value today at $T=0$, the two light states correspond to the Standard Model top and charm quarks while the three heavy ones are the exotic FN fermions.}
\label{fig:yukawas}
\end{figure}

To illustrate our scenario we focus on the top and charm quarks using the Froggatt-Nielsen (FN) mechanism. The mass matrix follows the pattern,
	\begin{equation}
	\frac{\phi}{\sqrt{2}} \begin{pmatrix} \bar{t_{R}} \\ \bar{c_{R}} \end{pmatrix}^{T} \begin{pmatrix} 1 & \epsilon^{2} \\ \epsilon & \epsilon^{3} \end{pmatrix} \begin{pmatrix} t_{L} \\ c_{L} \end{pmatrix},
	\end{equation}
where $\epsilon \sim 0.2$. In the FN picture $\epsilon \equiv a_{s}/\Lambda_{\rm FN}$ where $\Lambda_{\rm FN}$ is the FN scale set by the mass of vector-like quarks and $a_{s}$ is the flavon VEV or an explicit soft breaking of the FN symmetry by one unit. It is necessary to explicitly break the FN symmetry, unless the FN symmetry is gauged in an extended model, in order to avoid the appearance of a massless Goldstone boson. Hence, we shall assume below that $a_{s}$ arises from an explicit breaking, in order to avoid having to study the dynamics of the flavon field.\footnote{The flavon will eventually gain a VEV and --- if it is of the same order as the other dimensionful terms in the flavour sector --- it will also be of the same order as the explicit breaking scale. Hence it is not expected to change our overall picture. This means the scalar and pseudoscalar components of the flavon will also end up with masses at a similar scale. To be safe from limits from $K-\bar{K}$ mixing this mass scale should be at least several TeV if the flavon couples to all quark flavours~\cite{Baldes:2016gaf}.} To generate the above mass matrix we assign the following FN charges to the SM quarks:
	\begin{equation}
	Q_{\rm FN}\begin{pmatrix} t_{L} \\ b_{L} \end{pmatrix} = 0, \qquad Q_{\rm FN}\begin{pmatrix} c_{L} \\ s_{L} \end{pmatrix} = -2, \qquad Q_{\rm FN}(t_{R}) = 0, \qquad Q_{\rm FN}(c_{R}) = 1.
	\end{equation}
In the UV completion we add vector-like quarks which transform as $u_{R}$ under the SM gauge group. We require three such vector-like quarks, 
	\begin{equation}
	G_{L,R}^{0} \qquad G_{L,R}^{1} \qquad G_{L,R}^{2},
	\end{equation}
each with twelve degrees-of-freedom, where $L$ and $R$ denote the usual chiral components and the number in the superscript denotes the negative FN charge. The full mass matrix is then given by
	\begin{equation}
	\frac{1}{\sqrt{2}}\begin{pmatrix} \bar{G}_{R}^{0} \\\bar{G}_{R}^{1} \\ \bar{G}_{R}^{2}\\ \overline{t}_{R} \\ \overline{c}_{R} \end{pmatrix}^{T}
	 \begin{pmatrix} M & a_{s} & 0 & \phi & 0 \\
			 a_{s} & M & a_{s} & 0 & 0 \\
			 0 & a_{s} & M & 0 & \phi \\	
			 M & a_{s} & 0 & \phi & 0 \\
			 a_{s} & 0 & 0 & 0 & 0 
	 \end{pmatrix}
	\begin{pmatrix} G_{L}^{0} \\ G_{L}^{1} \\ G_{L}^{2} \\ t_{L} \\ c_{L} \end{pmatrix},
	\end{equation}
where we have suppressed factors of $\mathcal{O}(1)$ and indicate bare mass terms allowed by the FN and electroweak symmetries by $M$. The entries proportional to $a_{s}$ break the FN symmetry by one unit. In principle there may be even smaller entries in the terms which break the FN symmetry by more than one unit, which may be generated by renormalization group flow. For simplicity we assume these are negligible and set the corresponding entries to zero. We next imagine the mass terms $M$ as arising from a bare contribution, which we take to be $\sim a_{s}$, and through the Yukawa coupling to another scalar $\Delta$ in the form $\Delta \bar{G}_{R}G_{L}$. The full mass matrix is therefore given by
	\begin{equation}
	\label{eq:fullmassmatrix}
	\frac{1}{\sqrt{2}}\begin{pmatrix} \bar{G}_{R}^{0} \\\bar{G}_{R}^{1} \\ \bar{G}_{R}^{2}\\ \overline{t}_{R} \\ \overline{c}_{R} \end{pmatrix}^{T}
	 \begin{pmatrix} a_{s}+\Delta & a_{s} & 0 & \phi & 0 \\
			 a_{s} & a_{s}+\Delta & a_{s} & 0 & 0 \\
			 0 & a_{s} & a_{s}+\Delta & 0 & \phi \\	
			 a_{s}+\Delta & a_{s} & 0 & \phi & 0 \\
			 a_{s} & 0 & 0 & 0 & 0 
	 \end{pmatrix}
	\begin{pmatrix} G_{L}^{0} \\ G_{L}^{1} \\ G_{L}^{2} \\ t_{L} \\ c_{L} \end{pmatrix}.
	\end{equation}
The $\Delta$ eventually obtains a large VEV, $\langle \Delta \rangle \equiv v_{\Delta}$, giving $\epsilon \approx  a_{s}  / v_{\Delta} \approx 1/5$. It is useful to define effective Yukawa couplings
	\begin{align}
	y^{\rm eff}_{f \phi} = \sqrt{2} \frac{\partial m_{f}}{\partial \phi}, \qquad \qquad \qquad
	y^{\rm eff}_{f \Delta} = \sqrt{2} \frac{\partial m_{f}}{\partial \Delta}.
	\end{align}
The effective Yukawa couplings to the Higgs and the fermion masses along the $\phi$ axis are shown in Fig.~\ref{fig:yukawas}. At the zero temperature minimum (where $\Delta=v_{\Delta}$, not shown in in Fig.~\ref{fig:yukawas}), we obtain three super heavy mass eigenstates, $m_{f} \sim v_{\Delta}$, corresponding to the FN fermions, a mass eigenstate corresponding to the top, $m_{t} \sim v_{\phi}$, and one corresponding to the charm, $m_{c} \sim \epsilon^{3} v_{\phi}$. Below we shall study the temperature evolution of $\Delta$ together with the Higgs field $\phi$. The numerical values of the $\mathcal{O}(1)$ coefficients in the fermionic mass matrix used in our analysis are given in Appendix~\ref{Sec:coefficients}.

\begin{figure}[t]
\begin{center}
\includegraphics[width=200pt]{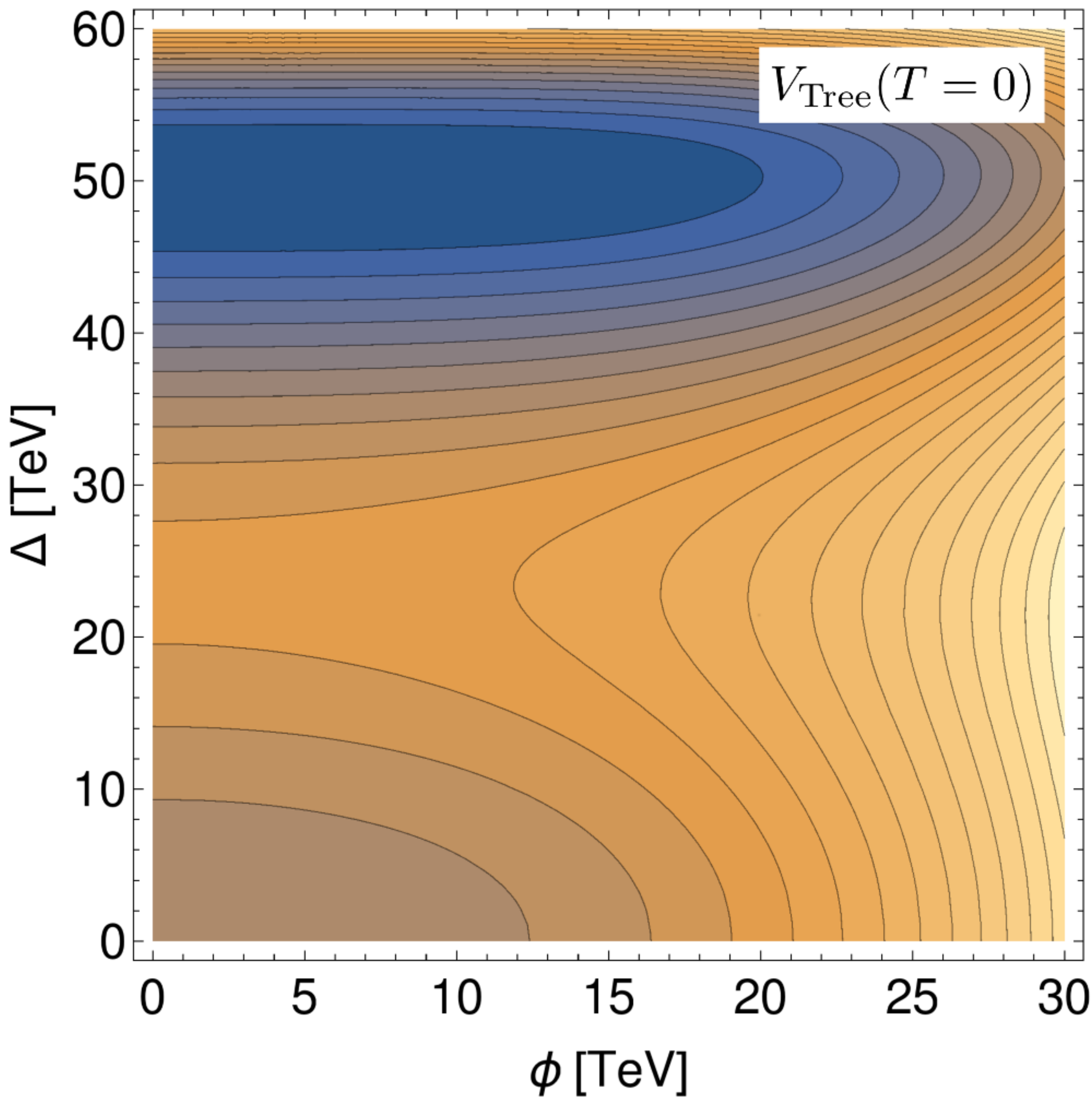} \qquad
\includegraphics[width=200pt]{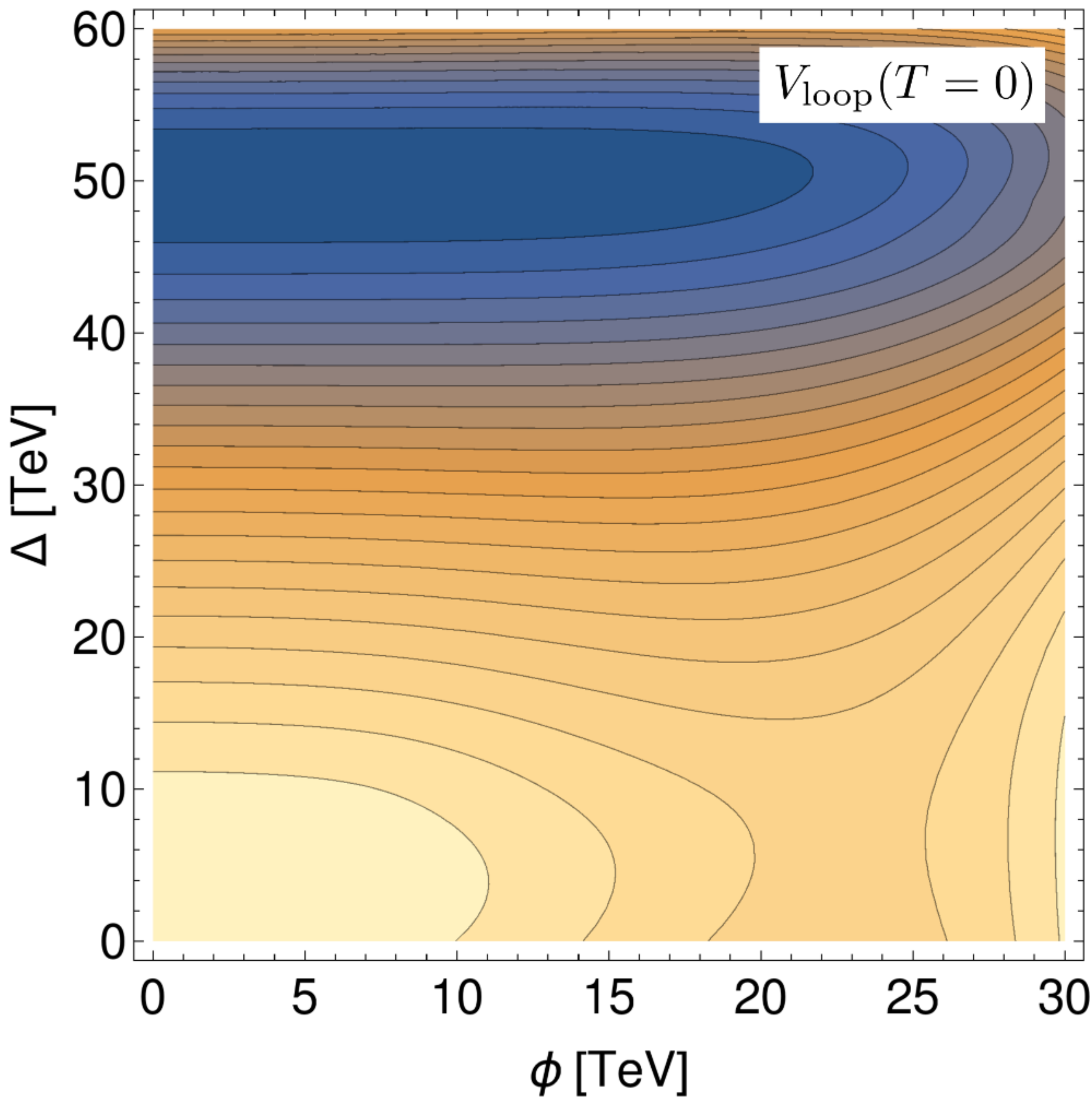}
\end{center}
\caption{\small The tree-level potential (left) and the one-loop zero temperature potential (right). The parameters have been chosen as given in Eqs.~(\ref{eq:exampleparameters}) and (\ref{eq:parameters}). Note the shift in the minimum along the $\phi$ axis due to the one-loop effect of the fermions.}
\label{fig:zeroTpot}
\end{figure}

\subsection{The scalar potential}
In this section we describe the two-field scalar potential consisting of the real scalar $\Delta$ and the electroweak Higgs $\phi$. We write the tree-level potential as
	\begin{align}
	V(\phi,\Delta)   = & \; \frac{\mu_{\phi}^{2}}{2}\phi^{2} + \frac{\lambda_{\phi}}{4} \phi^{4} + \frac{\lambda_{\phi \Delta}}{4}\phi^{2}\Delta^{2} +  \frac{\mu_{\Delta}^{2}}{2}\Delta^{2} + \frac{\lambda_{\Delta}}{4} \Delta^{4} \nonumber \\
			 &  \; + \frac{1}{8\Lambda_{a}^{2}}\Delta^{6} + \frac{1}{8\Lambda_{b}^{2}}\phi^{2}\Delta^{4} + \frac{1}{8\Lambda_{c}^{2}}\phi^{4}\Delta^{2} + \frac{1}{8\Lambda_{d}^{2}}\phi^{6}.
	\end{align}
We fix the electroweak Higgs mass and VEV to the observed values, $m_{h} = 125$ GeV and $v_{\phi} = 246$ GeV. The higher dimensional terms are required to stabilise the potential due to the large number of fermions with sizable Yukawa couplings. In particular $\Lambda_{a}$ should not be too far above the scale of the FN fermions or there would be an instability in the $\Delta$ direction. Note that generically, to achieve $m_{h} = 125$ GeV and $v_{\phi} = 246$ GeV requires a large degree of fine-tuning already in the tree level terms of the potential as soon as $v_{\Delta} \gg v_{\phi}$, which is required from flavour constraints. This is seen in the relation for the physical Higgs mass
	\begin{equation}
	\label{eq:higgsmassparameter}
	m_{h}^{2} \sim \mu_{\phi}^{2} + 3 \lambda_{\phi} v_{\phi}^{2} + \frac{\lambda_{\phi \Delta}}{2} v_{\Delta}^{2} + \frac{1}{4\Lambda_{b}^{2}}v_{\Delta}^{4} + \frac{3}{2\Lambda_{c}^{2}}v_{\Delta}^{2}v_{\phi}^{2}+ \frac{15}{4\Lambda_{d}^{2}}v_{\phi}^{4},
	\end{equation}
where the large terms on the right-hand-side must be tuned to return the much smaller $m_{h}^{2}$. Here our philosophy is to assume these large tree-level contributions cancel, possibly due to a Higgs relaxation mechanism operating during an earlier period, as discussed in Sec.~\ref{sec:relaxation}.
In our example we choose the parameters to be
	\begin{gather}
	 v_{\Delta} = 50 \; \mathrm{TeV}, \qquad \lambda_{\phi \Delta} = -0.05, \qquad \lambda_{\Delta} = -0.23, \nonumber \\  \Lambda_{a}  = \Lambda_{d}= 100 \; \mathrm{TeV}, \qquad  \Lambda_{b} = \Lambda_{c} = 300 \; \mathrm{TeV}.
	\label{eq:exampleparameters}
	\end{gather}
As we shall see below, the dimensionless couplings have been chosen as to obtain the required pattern of symmetry breaking in the thermal evolution of the potential. With the parameters chosen above we find $\lambda_{\phi} \approx 0.12$. The effective electroweak quartic, i.e. $\lambda_{\phi}+(v_{\Delta}/\sqrt{2}\Lambda_{c})^{2}+(v_{\phi}/\sqrt{2}\Lambda_{d})^{2}$, which enters the triple Higgs cross section for collider searches, remains close to its SM value. The potential is shown on the left in Fig.~\ref{fig:zeroTpot}. The one-loop potential, including the effect of the fermions after diagonalising~(\ref{eq:fullmassmatrix}), is shown on the right in Fig.~\ref{fig:zeroTpot}. As can be seen in the figure, although the tree-level potential has a barrier in the $\Delta$ direction, this is almost completely erased at loop-level once the effect of the fermions is taken into account. The appropriate strong first order phase transition can be achieved by an interplay of the fermionic degrees-of-freedom together with the symmetry non-restoration sector which we discuss next.

\begin{figure}[t]
\begin{center}
\includegraphics[width=220pt]{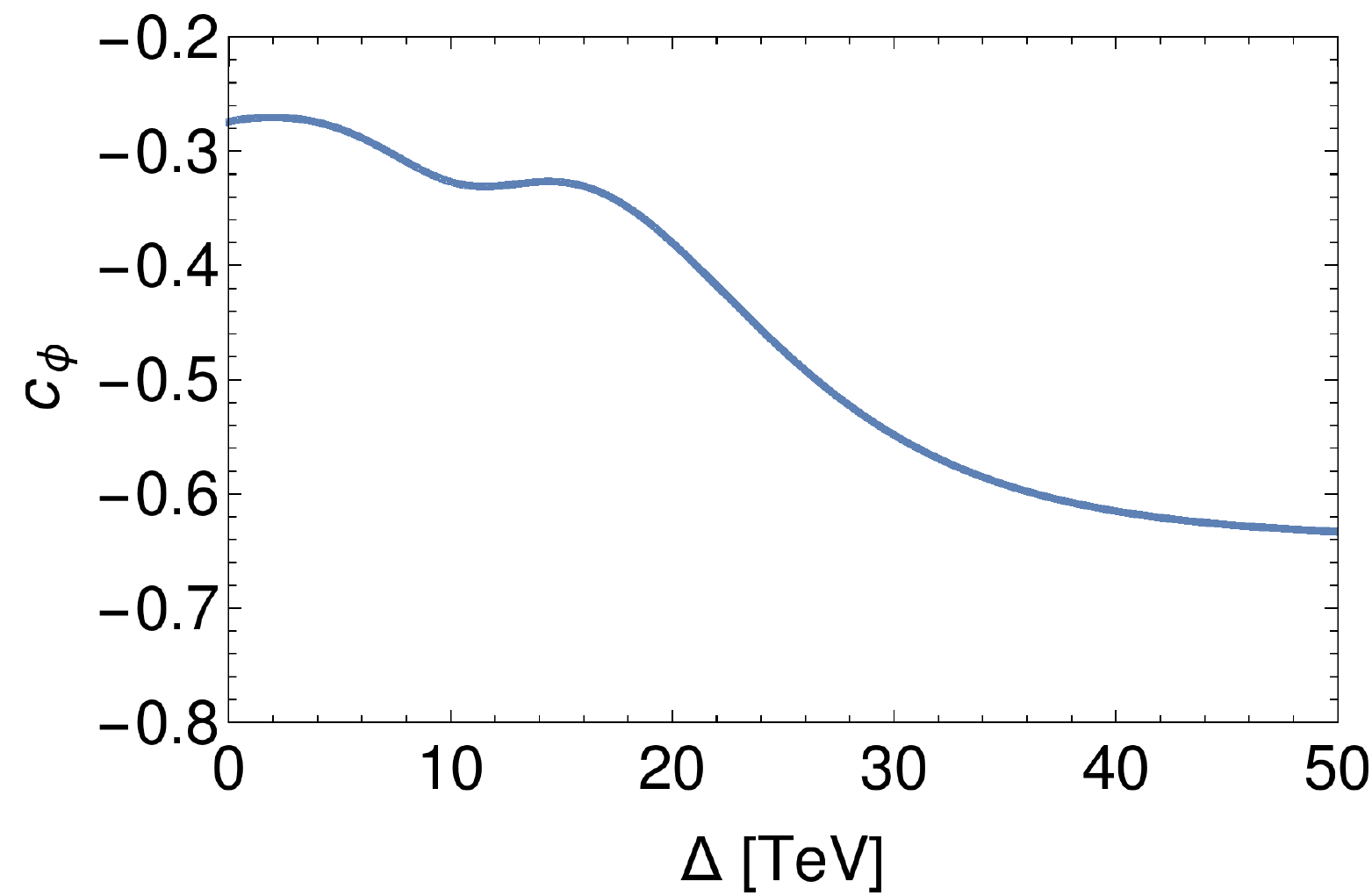}
\end{center}
\caption{\small The thermal mass coefficient of the Higgs at high $T$, above the relevant mass scales, for the parameters of Eq.~(\ref{eq:parameters}), with the Yukawas evaluated at $\phi = 20$ TeV.}
\label{fig:thermalmass}
\end{figure}

\subsection{Symmetry non-restoration sector}

As in the toy example, we obtain a negative contribution to the thermal mass of $\phi$ through negative cross quartics. Let us introduce $N_{\rm Gen}$ generations of  singlet scalars $\chi_{i}$, i.e.~each with $N_{\chi_{i}}=1$ degree of freedom. The relevant terms in the tree level potential are
	\begin{equation}
	V(\phi,\chi) = \frac{\lambda_{\phi \chi}}{4}\phi^{2}\sum_{i=1}^{N_{\rm Gen}}\chi_{i}^{2} +  \frac{\mu_{\chi}^{2}}{2}\sum_{i=1}^{N_{\rm Gen}}\chi_{i}^{2} + \frac{\lambda_{\chi}}{4}\sum_{i=1}^{N_{\rm Gen}}\chi_{i}^{4},
	\label{eq:phichipot}
	\end{equation}
where we again assume for simplicity that the couplings are universal and that any interactions between the $\chi_{i}$ and $\Delta$ are negligible (also in order to keep $c_{\chi}$ small enough to not spoil the symmetry non-restoration effect). At high temperatures the thermal masses of the fields are
	\begin{gather}
	c_{\chi_{i}}T^{2}  \approx \left( \frac{\lambda_\chi}{4} + \frac{ \lambda_{\phi\chi} }{ 6 } \right)T^{2}, \\
	c_{\phi}T^{2}  \approx \left( \frac{\lambda_{\phi}} {2} + 3\frac{g_{2}^2}{16} + \frac{g_{Y}^2}{16}  + \frac{\lambda_{\phi \Delta}}{24} + N_{\rm Gen} \frac{\lambda_{\phi \chi}}{24} + \frac{1}{ 4 }\sum_{f} \left[y_{f \phi}^{\rm eff}(\phi,\Delta)\right]^{2}  \right)T^{2}, \label{phithermalmass}
	\end{gather}
where we have introduced the sum over the effective, field-dependent, Yukawa couplings of the Higgs. As the SM contributions to $c_{\phi}$ already amount to $\approx 0.4$, we require a large $N_{\rm Gen}$ in order to obtain a negative thermal mass for $\phi$ while remaining consistent with the stability constraint.\footnote{The stability constraint is relaxed in the presence of the higher dimensional operators. Here we choose parameters sufficient for stability of the potential, i.e. consistent with stability in the limit $\Lambda_{b}, \;\Lambda_{c},\;\Lambda_{d} \to \infty$.} Here, for illustration, we choose
	\begin{equation}
	N_{\rm Gen}=2000, \qquad N_{\chi i}=1, \qquad \lambda_{\chi}=0.7, \qquad \lambda_{\phi \chi}=-0.012.
	\label{eq:parameters}
	\end{equation}
Note that the size of the symmetry non-restoration effect depends on the field values $\phi$ and $\Delta$, through the effective Yukawas $y_{\phi i}^{\rm eff}$. This is illustrated in Fig.~\ref{fig:thermalmass}. Nevertheless, this does not mean there is necessarily a minimum for non-zero $\phi$ and $\Delta$ at all temperatures, because of (i) the higher dimensional terms and because (ii) the FN fermions also couple to $\Delta$ raising the potential due to finite temperature effects. Because some of the fermions already have masses of the order of the critical temperature in the symmetric phase,  $m_{f} \sim a_{s} \sim T_{c}$, a simple use of Eq.~(\ref{phithermalmass}) is not possible here, and a numerical evaluation of the effective potential is required.  

\begin{figure}[t]
\begin{center}
\includegraphics[width=220pt]{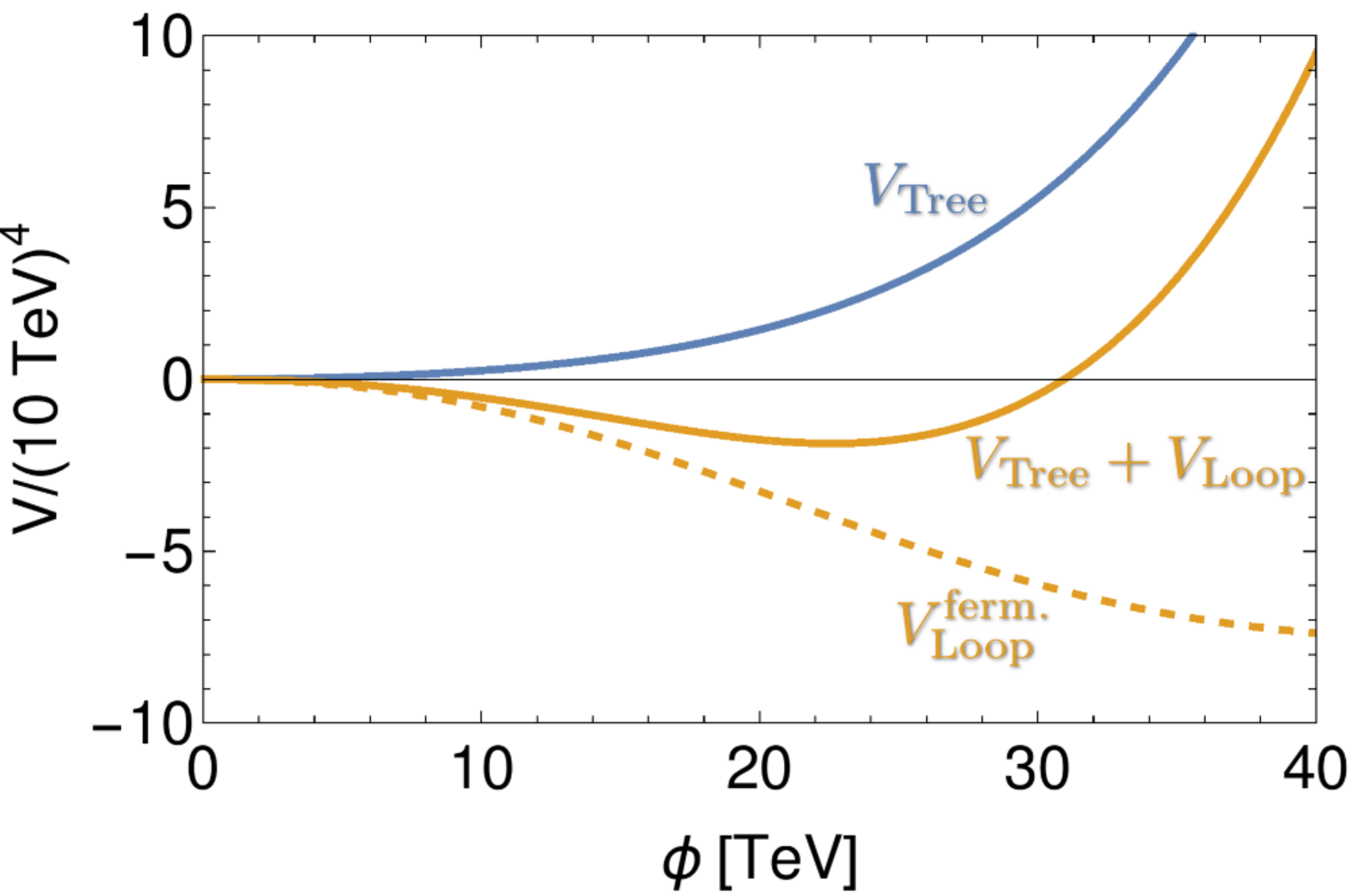}
\includegraphics[width=220pt]{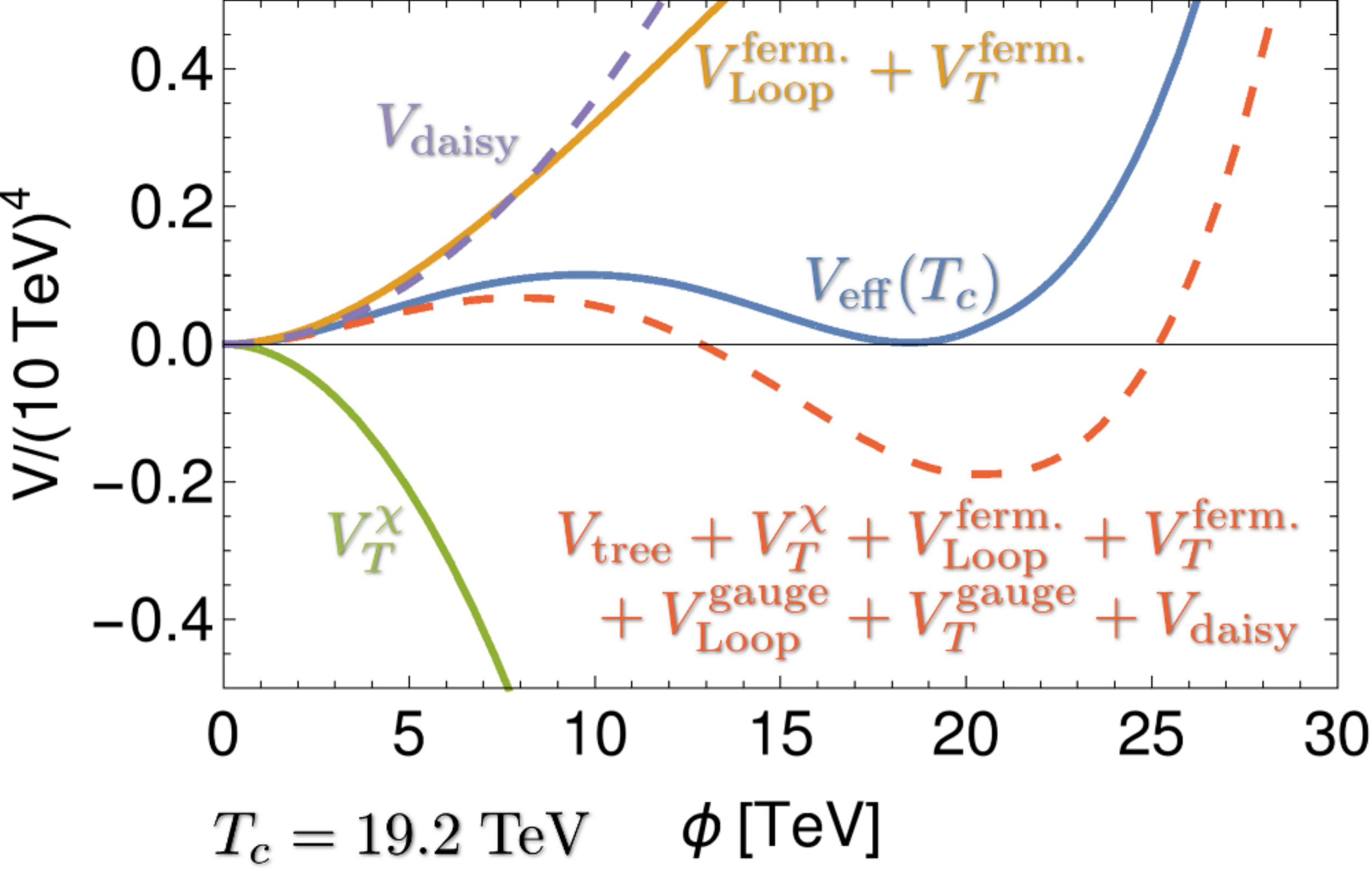}
\end{center}
\caption{\small Left: The tree-level and one-loop effective potential along the $\phi$ axis at $T=0$ showing the role of the fermions in shifting and deepening the broken phase minimum. Right: The effective potential (blue line) at the EW phase transition critical temperature $T_{c}=19.2$~TeV. The green line shows the thermal contribution of the scalar $\chi$. The yellow line shows the $T=0$ and thermal contribution from the  fermions. The dashed purple line is the daisy contribution. The red dashed line shows the sum of the tree level potential, the $T=0$ and thermal contribution of the gauge bosons, and the other contributions above, showing these terms lead to a barrier.}
\label{fig:barrier}
\end{figure}

\begin{figure}[p!]
\begin{center} 
\includegraphics[width=200pt]{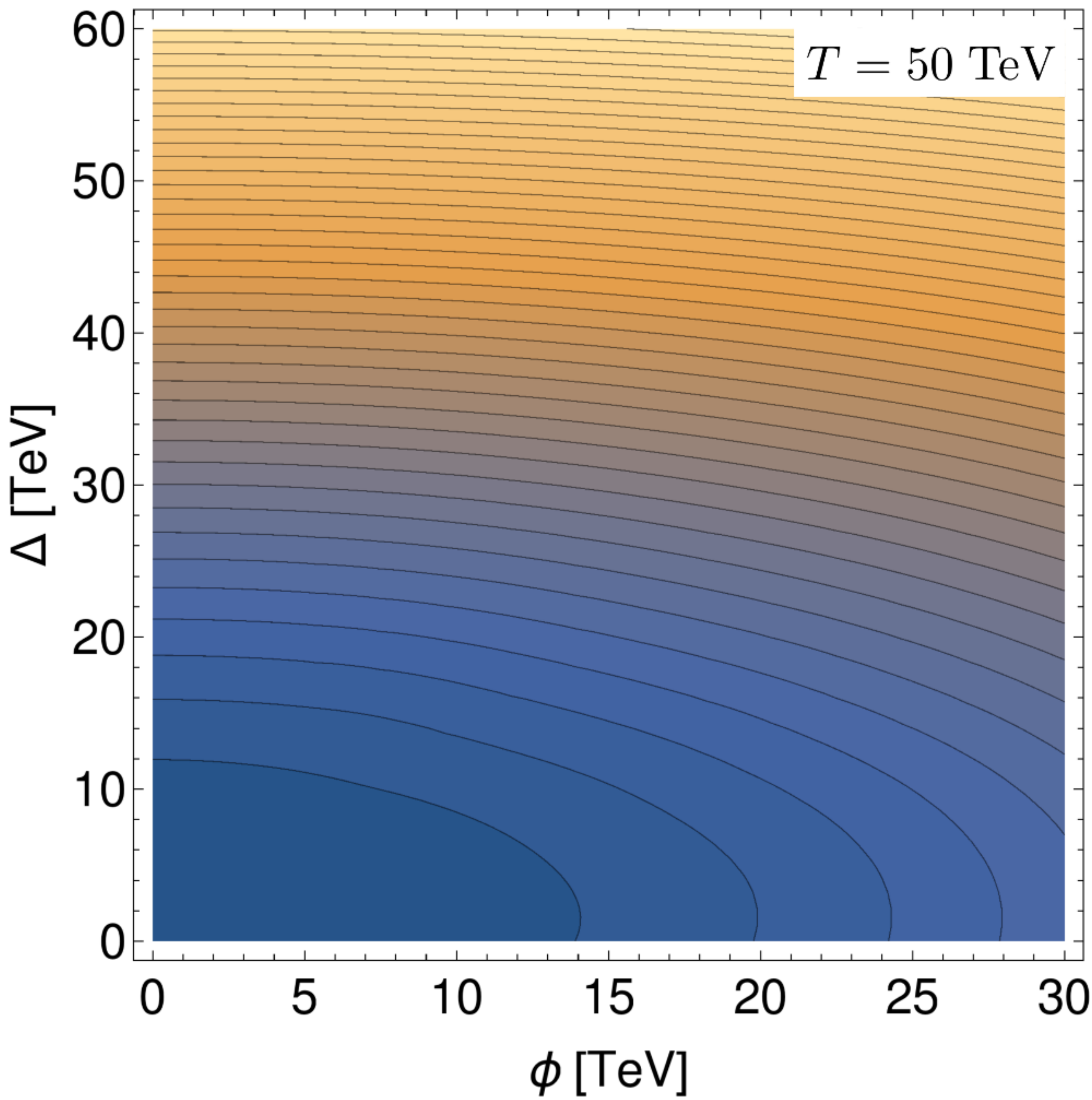} \qquad
\includegraphics[width=200pt]{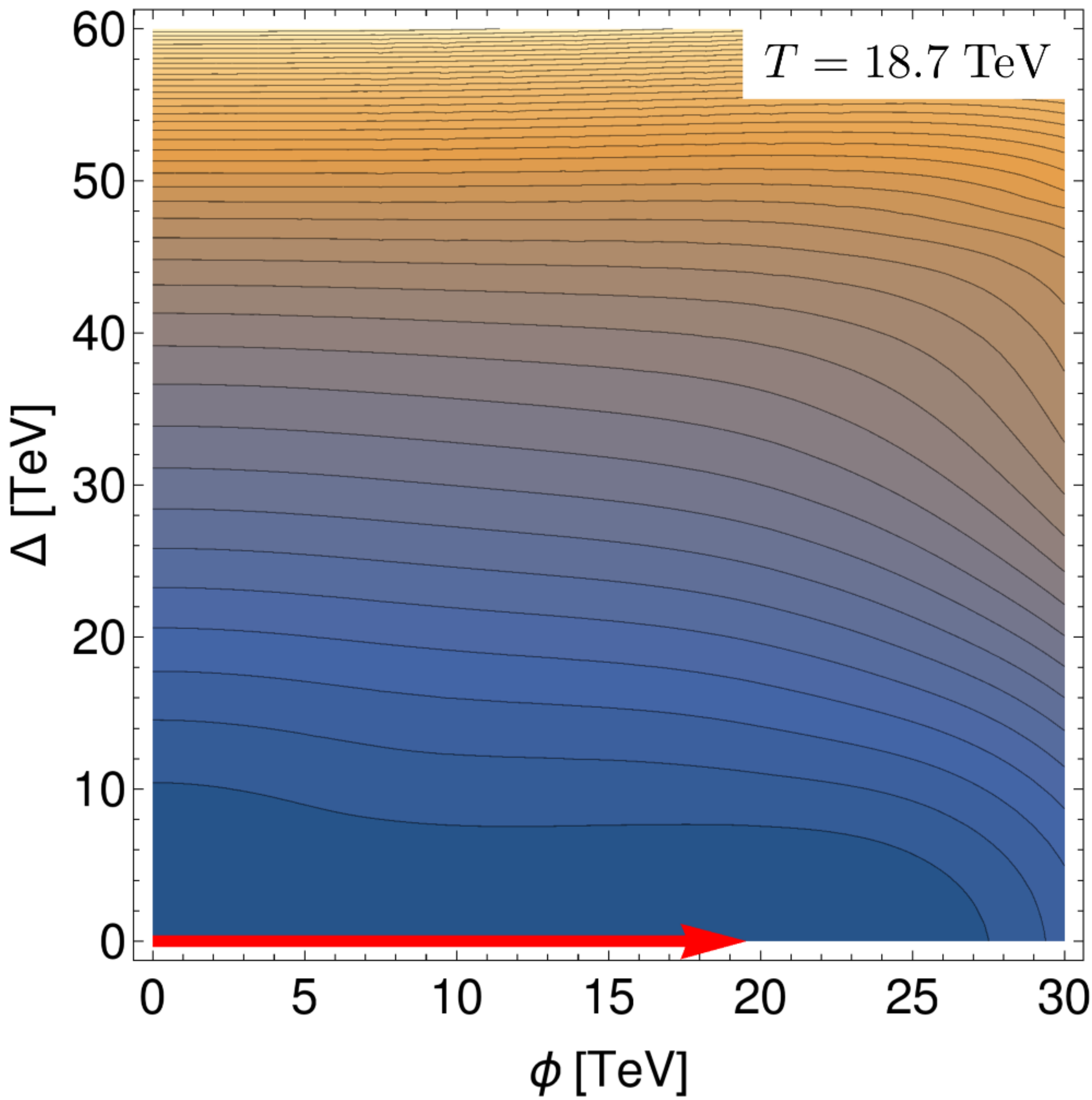} \\
\vskip 8pt
\includegraphics[width=200pt]{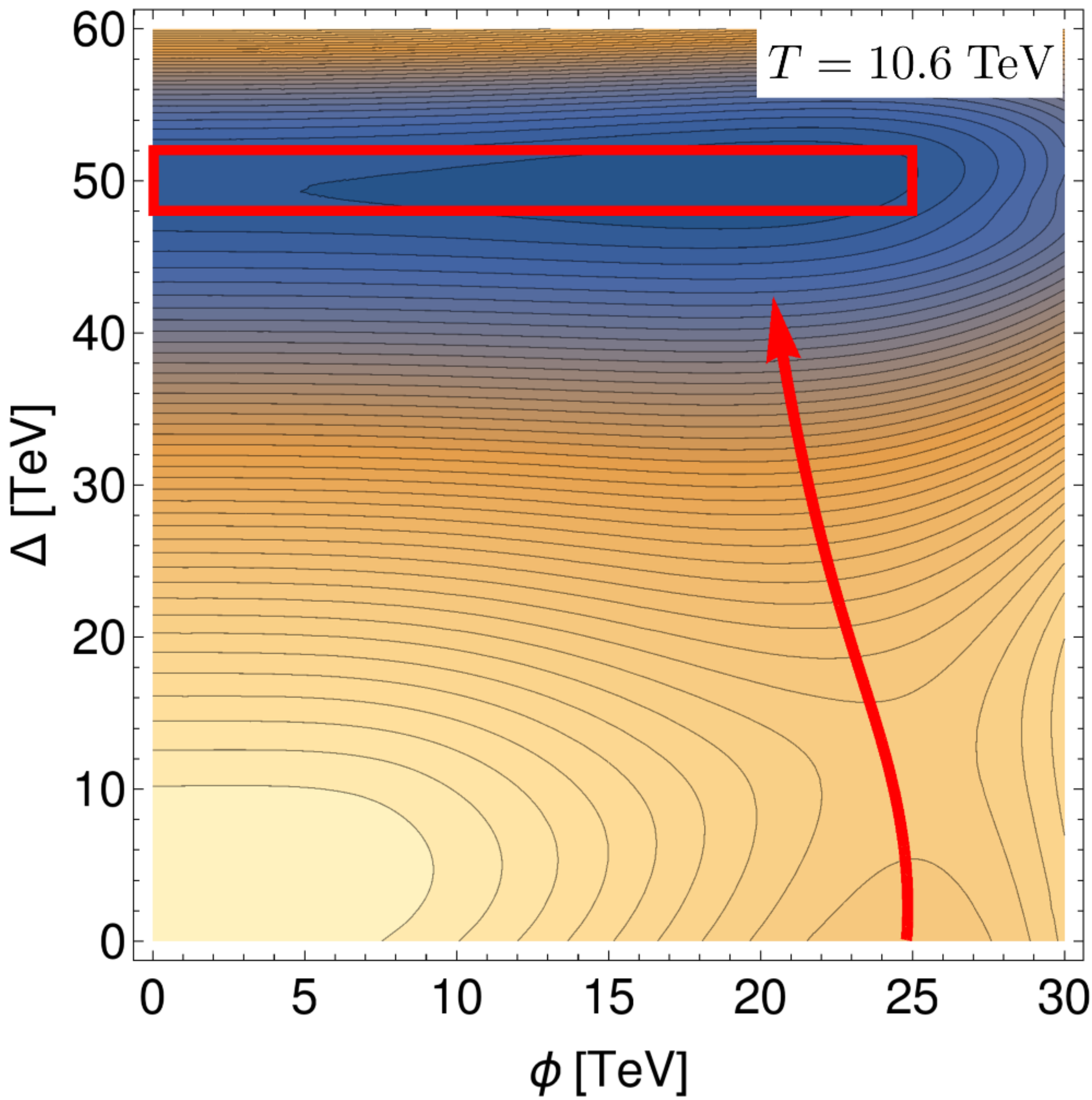} \qquad
\includegraphics[width=200pt]{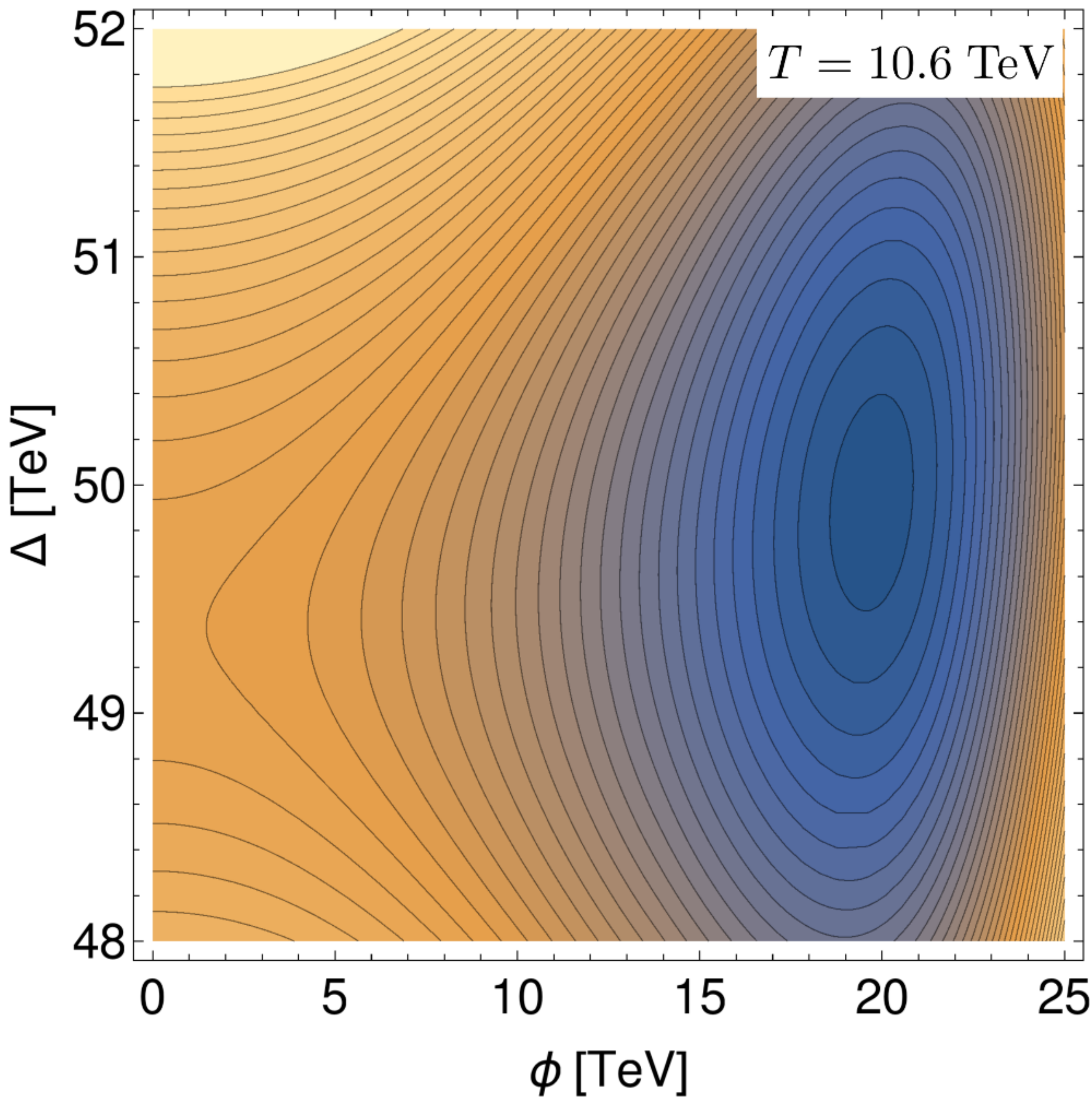}
\end{center}
\caption{\small Evolution of the effective potential. 
{\bf Step 0 (top left)}: in the symmetric phase at $T=50$ TeV. 
{\bf Step 1 (top right)}: first order phase transition in the $\phi$ direction at $T_{n}\approx 19$ TeV. The path of the bounce solution when $S_{3}/T \approx 123$ is shown as a red line. 
{\bf Step 2 (bottom left)}: First order phase transition mostly in the $\Delta$ direction at $T_{n}\approx 11$ TeV. The rectangular area corresponds area in the bottom right plot. 
{\bf Step 3 (bottom right)}: Magnification of a selected area of the potential at $T \approx 11$ TeV, showing the presence of the EW minimum with $\phi/T \gtrsim 1$. This minimum subsequently decreases with temperature to the present day value $v_{\phi}=246$ GeV.
}
\label{fig:potevo}
\pagestyle{empty}
\end{figure}

\section{Phase transition and evolution of the potential}
\label{sec:phasetrans}

We now describe the various effects contributing to the sequence of phase transitions in this model. As advertised above, temperature effects from the additional FN fermions play a crucial role in maintaining the global minimum of the potential at $\phi=0$ and $\Delta=0$ at high temperatures. Eventually $T$ drops, a non-trivial interplay between the one-loop terms for the fermions, assisted by those for the EW gauge bosons, and the symmetry non-restoration effect allows for a broken phase minimum to develop. Depending on the parameters chosen, we find we can obtain a first order phase transition along the $\phi$ direction, followed by another first order phase transition in the $\Delta$ direction.
Afterwards there is a slow evolution to the zero temperature minimum, at large $\Delta$ and small $\phi$, all the while maintaining $\phi/T \gtrsim 1$.

In order to avoid an early transition along the $\Delta$ direction, which leads eventually to a cross over in the $\phi$ direction due to the negative thermal mass at large $\Delta$ values, the quartic $\lambda_{\Delta}$ should be small enough.
The mass parameter $\mu_{\chi}$ must be at most EW scale, in order to maintain a negative $c_{\phi}$ down to $T \sim v_{\phi}$, for simplicity we have set it to zero for our plots in this section. We discuss relaxing this assumption in Sec.~\ref{sec:IR}.
In order to obtain a strong first order phase transition in the $\phi$ direction, we require $N_{\rm Gen}|\lambda_{\phi \chi}|$ to be below some value, otherwise the phase transition occurs too early. On the other hand, to maintain a large enough $\phi/T$ during the subsequent evolution, we require a large enough $N_{\rm Gen}|\lambda_{\phi \chi}|$. Keeping all other parameters fixed, we find the correct evolution of the potential for $1500 \lesssim N_{\rm Gen} \lesssim 2000$, when we set $\mu_{\chi} = 200$ GeV (which is relevant for the low $T$ analysis).

A detailed plot of the potential at the critical temperature is shown in Fig.~\ref{fig:barrier}. Note the interplay between the symmetry non-restoration effect --- arising from loop effects of $\chi$ on the potential --- and the positive fermionic, gauge and daisy terms leads to the barrier. We wish to emphasise the particular importance of the \emph{varying} Yukawas in achieving a strong first order phase transition in our example~\cite{Baldes:2016rqn}. The phase transition, in contrast, is much weaker if we switch off the Yukawa variation effects. This is discussed in further detail in Appendix~\ref{Sec:constyuk}. The overall evolution of the potential for our choice of parameters is shown in Fig.~\ref{fig:potevo}. We have calculated the $O(3)$ symmetric bounce action for the bubble, denoted $S_{3}$, using the AnyBubble code~\cite{Masoumi:2016wot}. The probability of nucleating a bubble in a Hubble volume reaches $\sim 1$ in a radiation dominated Universe when~\cite{Moreno:1998bq}
	\begin{equation}
	\frac{S_{3}}{T} \approx 4 \; \mathrm{ln}\left( \sqrt{ \frac{45}{4\pi^{3}g_{\ast}} } \frac{M_{\rm Pl}}{T} \right) \approx 123 - 4 \; \mathrm{ln}\left( \frac{T}{10 \; \mathrm{TeV} } \right) - 2 \; \mathrm{ln}\left(\frac{g_{\ast}}{1000}\right),
	\end{equation}
where $M_{\rm Pl}$ is the Planck mass and $g_{\ast}$ counts the effective radiation degrees-of-freedom contributing to the Hubble expansion~\cite{Drees:2015exa,Borsanyi:2016ksw}, which now includes the $\chi_{i}$ contribution.

In our example we find the step 1 phase transition occurs at $T_{n} \approx 19$ TeV. Here the washout parameter reads
	\begin{equation}
	\frac{\phi_{n}}{T_{n}} \approx 1.0.
	\end{equation}
After remaining in the $\phi \sim 20$ TeV minimum and supercooling to $T \approx 11$ TeV, we find the step 2 phase transition occurs. The path of the two-field bounce solution is shown in Fig.~\ref{fig:potevo}. Note the initial bubble is thick-walled, meaning the centre of the bubble is away from the true minimum in field space. Nevertheless, as the bubble expands the fields will quickly relax down to the minimum of the potential. We have checked that with the given parameters the minimum does indeed respect $\phi/T \gtrsim 1$ until the EW minimum is reached. We have also checked that the positive thermal contributions in the $\chi_{i}$ field direction are sufficient to keep the $\chi_{i}$ VEVs at zero throughout.

\section{Earlier cosmological history}
\label{sec:relaxation}

We now comment on the hierarchy problem in this framework.
As shown in Eq.~(\ref{eq:higgsmassparameter}), we need to tune parameters to keep the Higgs mass parameter $m_h^2$ small. One way to address this is to stipulate that a relaxion mechanism took place before the EW phase transition~\cite{Graham:2015cka}. The relaxation of the Higgs parameter would have to take place during inflation.
We would then have the following cosmological history:
\begin{enumerate} 

\item  Inflation begins, drastically lowering the temperature of the thermal bath. At this stage, the Higgs mass parameter and the $\Delta$ mass parameter are both large.

\item  Relaxation starts for the Higgs and its mass parameter, Eq.~(\ref{eq:higgsmassparameter}), is relaxed to the usual low weak scale value. Relaxation ends when the Higgs obtains a small VEV, leading to the backreaction on the relaxion potential. The VEV of $\Delta$ does not contribute to the relaxion potential barriers and hence it can naturally be of a larger scale.

\item  Reheating in the visible sector: EW symmetry is restored and $\Delta$ is also reheated so its VEV goes back to zero.

\item The evolution described in Sec.~\ref{sec:phasetrans} takes place.

\end{enumerate}

In this context, we would assume that the $\chi$ scalar sector is also relaxed during stage 3.
So the Higgs and $\chi$ are part of a common sector (perhaps composite) and they are relaxed together, while $\Delta$ from the flavour sector is not subject to relaxation.

\section{Phenomenology}
\label{sec:pheno}

\subsection{Gravitational wave signal}

\begin{figure}[t]
\begin{center}
\includegraphics[width=260pt]{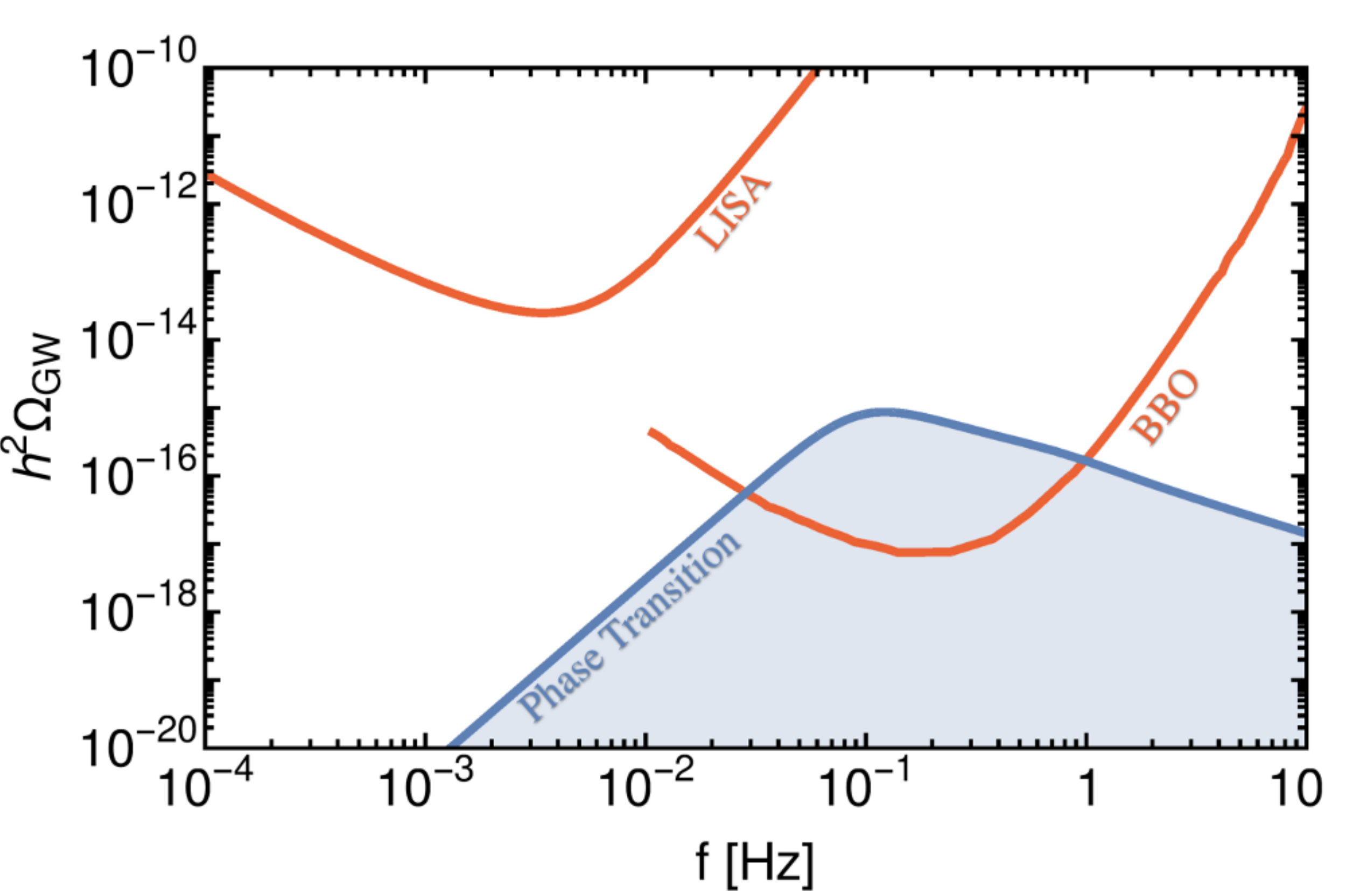}
\end{center}
\caption{\small The stochastic gravitational wave background generated by the step-2 phase transition, occurring at $T\sim 11$ TeV, assuming a bubble wall velocity $v_{w}=1$, compared with prospects for future gravitational wave observatories.
}
\label{fig:GWspectrum}
\end{figure}

During the cosmological evolution,  after the EW phase transition and baryogenesis, between step 1 and step 2, the scalar fields become stuck at a false minimum and there is some super cooling. The timescale of the transition is
	\begin{equation}
	\frac{\beta}{H} \equiv T_{n} \frac{d}{dT}\left(\frac{S_3}{T}\right)\bigg|_{T_n} \approx 180.
	\end{equation}	 
The ratio of energy released compared to radiation bath, however, is rather suppressed
	\begin{equation}
	\alpha \equiv \frac{ \rho_{\rm vac}(\mathrm{false})- \rho_{\rm vac}(\mathrm{true}) }{ \rho_{\rm rad} } \approx 8 \times 10^{-3},
	\end{equation}
due to the contribution of $N_{\rm Gen}$ to $g_{ \ast}$. 
Thus the resulting stochastic gravitational wave background generated during the first-order phase 
transition~\cite{Grojean:2006bp} is suppressed.
It is  too small to be detected by LISA~\cite{Caprini:2015zlo} but it is within the BBO sensitivity~\cite{Thrane:2013oya}, as illustrated in Fig.~\ref{fig:GWspectrum}. Due to the unusual situation of a relatively low $\beta/H$ combined with a suppressed $\alpha$, the bubble wall collisions (envelope contribution), gives the dominant effect. The step-1 phase transition is characterised by $\beta/H \approx 7300$ and $\alpha \approx 10^{-5}$, and returns a completely negligible gravitational wave background.

On the other hand, if another cosmological gravitational wave background due to cosmic strings exists, the rapid and huge change in $g_{\ast}$ due to the $\chi_{i}$ at the EW scale leads 
to some feature at a characteristic frequency in the spectrum of gravitational waves emitted in the radiation dominated era~\cite{Cui:2017ufi}. A similar feature can be expected in suitable gravitational wave backgrounds coming from inflation. We leave this study for future investigation.

\subsection{Scalar sector in the IR}

\begin{figure}[t]
\begin{center}
\includegraphics[width=220pt]{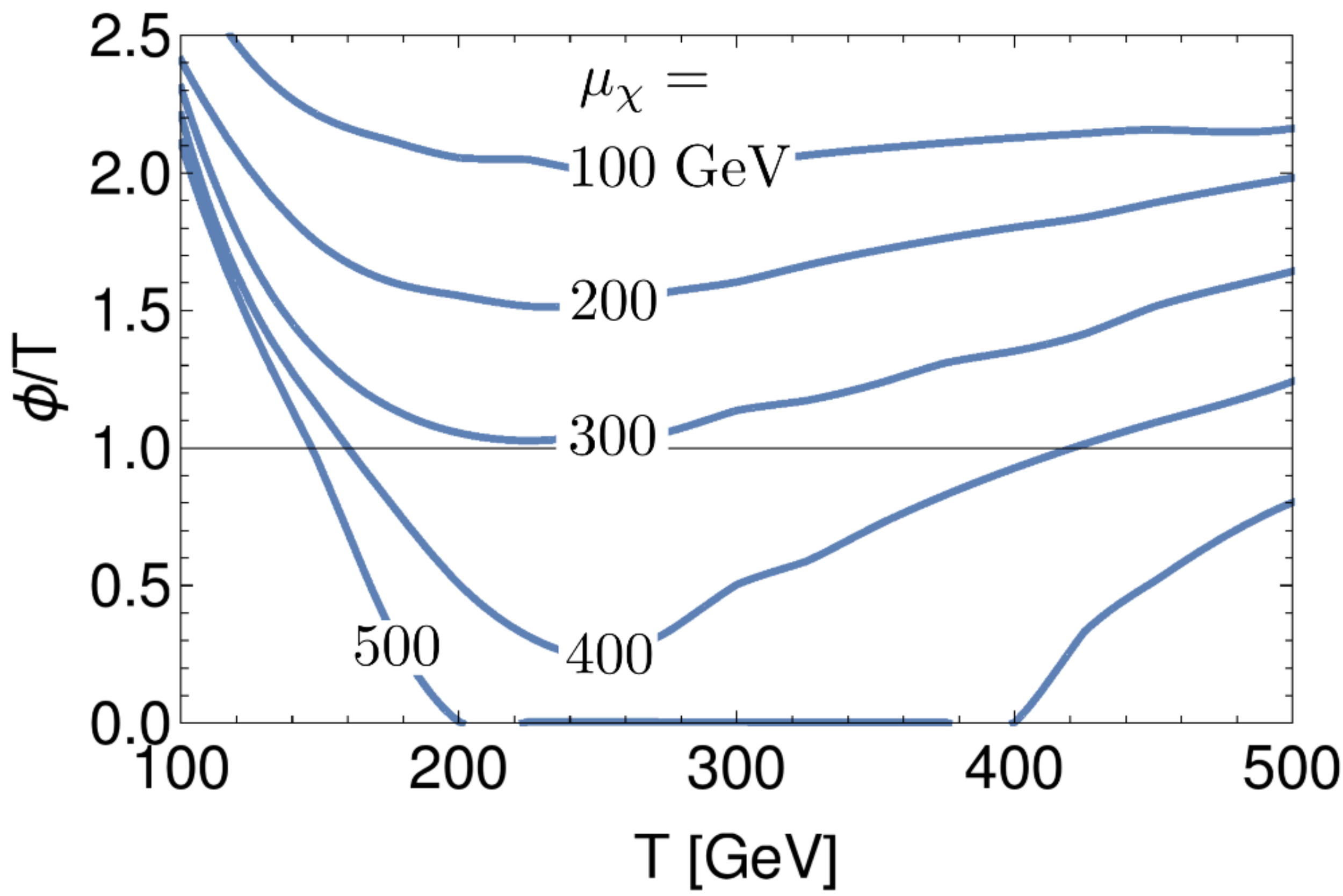}
\end{center}
\caption{\small The evolution of $\phi/T$ for different values of $\mu_{\chi}$. To retain $\phi/T\gtrsim 1$ we require $\mu_{\chi} \lesssim 300$ GeV.}
\label{fig:phioverT}
\end{figure}

\label{sec:IR}

The scenario relies on the scalar degrees-of-freedom $\chi_{i}$ to guide the electroweak minimum to its present value. Hence, it is necessary for the mass $\mu_{\chi}$ to be at or below the EW scale otherwise, once $T \lesssim \mu_{\chi}$, the symmetry non-restoration effect disappears and $\phi/T$ will become small. This is shown in Fig.~\ref{fig:phioverT}. The experimental constraint on such a scenario comes from searches for these light scalars. Note while we have considered universal mass and coupling terms for the $\chi_{i}$, we can imagine that in a more realistic scenario the masses are split in a spectrum of states with masses $m_{\chi_i}^{2} \sim \mathcal{O}(\mu_{\chi}^{2}+\lambda_{\phi \chi}v_{\phi}^{2}/2)$. The partial width of the SM Higgs to the $\chi_{i}$ is given by
	\begin{equation}
	\sum_{i}\Gamma(\phi \to \chi_{i}\chi_{i}) = \sum_{i} \frac{ \lambda_{\phi \chi}^{2}v_{\phi}^{2} }{ 32\pi m_{\phi} } \mathrm{Re} \left [ \sqrt{ 1- 4\frac{m_{\chi_i}^{2} }{ m_{\phi}^{2} } } \, \right] \sim N_{\rm Gen}' \frac{ \lambda_{\phi \chi}^{2}v_{\phi}^{2} }{ 32\pi m_{\phi} },
	\end{equation}
where $N_{\rm Gen}'$ denotes the number of generations with mass below the threshold $2m_{\chi_i} < m_{\phi}$. Demanding at most an $\mathcal{O}(0.1)$ modification to the SM Higgs signal strength requires $N_{\rm Gen}' \lesssim \mathcal{O}(1)$ for $\lambda_{\phi \chi} \sim 10^{-2}$. Hence the states must lie above this threshold. In summary, we obtain
	\begin{equation}
	63 \; \mathrm{GeV} \lesssim m_{\chi_{i}}  \lesssim 300 \; \mathrm{GeV},
	\end{equation}
by combining the EW Higgs decay constraint with the washout avoidance condition shown in Fig.~\ref{fig:phioverT}.

The $\chi_{i}$ states will become thermally populated and should not over-produce DM.  The cross quartic is too small for annihilation solely through the Higgs portal and anyway, at these masses, is ruled out by direct detection~\cite{Aprile:2017iyp,Cui:2017nnn,Akerib:2016vxi,Baldes:2013eva,Duerr:2015aka}. Hence we need to arrange for the $\chi_{i}$ to decay.\footnote{Alternatively, provided the additional interaction does not lead to a too large thermal mass, the $\chi_{i}$ could annihilate into dark radiation~\cite{Baldes:2017gzw}, or a dark mediator which subsequently decays~\cite{Baldes:2017gzu}.} This can be achieved if the $\chi_{i}$ obtain VEVs and can mix with the Higgs. Here we assume the $\chi_{i}$ obtain VEVs by introducing a small explicit breaking of the $Z_{2}$ symmetry $\chi_{i} \to -\chi_{i}$ in Eq.~(\ref{eq:phichipot}). Explicitly this may be introduced through a linear term in the scalar potential
	\begin{equation}
	V \supset  -\sum_{i} a_{\chi_{i}}^{3} \chi_{i},
	\end{equation}
which, in the limit $a_{\chi_{i}} \ll m_{\chi_{i}}$ induces VEVs
	\begin{equation}
	v_{\chi_{i}}  \sim \frac{ a_{\chi_{i}}^{3} }{ m_{\chi_{i}}^{2} }.
	\label{eq:chivev}
	\end{equation}
The mixing angle for the mixing of a singlet state with the Higgs is given by
	\begin{equation}
	\theta_{i} \approx \frac{\lambda_{\phi \chi}v_{\phi}v_{\chi_{i}}}{m_{\chi_{i}}^{2}-m_{\phi}^{2}}.
	\end{equation}
The $\chi_{i}$ can then mix with the SM Higgs and decay into light SM degrees of freedom. We demand that the $\chi_{i}$ decay before their energy density grows to dominate the universe, as otherwise they would dilute the baryon asymmetry~\cite{Scherrer:1984fd}. This can be achieved provided the decay rate of the $\chi_{i}$ states, $\Gamma_{i} \sim \theta_{i}^{2} \times 1$ MeV~\cite{Ellis:1975ap}, is larger than the Hubble rate, $H \sim \sqrt{g_{\ast}}T^{2}/M_{\rm Pl}$, when $T \sim m_{\chi i}$. This implies	$\theta_{i} \gtrsim 10^{-6}$ for $m_{\chi i} \sim 100$ GeV.

Further limits come from EW precision observables, Higgs signal strength measurements and direct searches~\cite{Robens:2015gla,Falkowski:2015iwa}. We may derive an approximate constraint by considering a degenerate spectrum, keeping in mind direct search limits will not apply once the masses are split in a more realistic model. Given a mass of the singlet states $m_{\chi_{i}} \sim 100$ GeV, the limit on the sum of the mixing angles reads $\sum_{i}|\theta_{i}| \lesssim 0.2 - 0.4$, depending on the mass~\cite{Robens:2015gla,Falkowski:2015iwa}. Together with the rapid decay condition, this bounds the mixing angle to lie in the range
	\begin{equation}
	10^{-6} \lesssim |\theta_{i}| \lesssim 10^{-4} \left( \frac{ 2000 }{ N_{\rm Gen} } \right).
	\end{equation}
Translated into a bound on the VEVs this reads
	\begin{equation}
	 \left( \frac{0.012 }{ |\lambda_{\phi\chi}| }\right) 5 \; \mathrm{MeV} \lesssim v_{\chi_{i}} \lesssim  \left( \frac{ 2000 \times 0.012 }{ |N_{\chi}\lambda_{\phi\chi}| }\right)  \mathrm{GeV},
	\end{equation} 
which shows Eq.~(\ref{eq:chivev}) can be applied consistently. 
By introducing a mixing with the SM Higgs, we also open up a decay channels of the form $\phi \to \chi_{i}^{\ast}\chi_{i} \to \bar{b}b\chi_{i}$, $\phi \to \phi^{\ast}\chi_{i} \to \bar{b}b\chi_{i}$, and $\phi \to \chi_{i}^{\ast}\chi_{i}^{\ast} \to \bar{b}b\bar{b}b$. Nevertheless, a calculation reveals that these are completely negligible. For example, from dimensional analysis,
	\begin{equation}
	\sum_{i}\Gamma(\phi \to \chi_{i}^{\ast}\chi_{i} \to \bar{b}b\chi_{i}) \sim \frac{3  N_{\rm Gen} \lambda_{\phi\chi}^{2}  \theta_{i}^{2} m_{b}^{2} }{ 128\pi^{3} m_{\phi} } 
		 \sim 10^{-10}\; \mathrm{MeV} \left( \frac{ N_{\rm Gen} }{ 2000 } \right) \left( \frac{ \lambda_{\phi\chi} }{ 0.012 } \right)^{2} \left( \frac{ \theta_{i} }{ 10^{-4} } \right)^{2}.
	\end{equation}
A similar calculation reveals
	\begin{equation}
	\sum_{i}\Gamma(\phi \to \phi^{\ast}\chi_{i} \to \bar{b}b\chi_{i})
	  \sim \frac{ 3  N_{\rm Gen} \lambda_{\phi\chi}^{2}  v_{\chi_{i}}^{2} m_{b}^{2} }{ 128\pi^{3}  v_{\phi}^{2} m_{\phi} } 
		  \sim 10^{-7}\; \mathrm{MeV} \left( \frac{ N_{\rm Gen} }{ 2000 } \right) \left( \frac{ \lambda_{\phi\chi} }{ 0.012 } \right)^{2} \bigg( \frac{ v_{\chi_{i}} }{ 1 \; \mathrm{GeV} } \bigg)^{2}.
	\end{equation}
We have confirmed these with a more detailed calculation, which also displays the additional suppression expected as $m_{\chi_{i}} \to m_{\phi}$ and the available phase space is reduced. The four-body decay channels are even more suppressed.

\section{Conclusions}

It is usually thought that the EW phase transition occurs when the Universe cools to temperatures $T \sim 100$ GeV. In this paper we have instead speculated on the possibility of high scale EW phase transition and EW baryogenesis.
This requires additional field content in order to break the EW symmetry at a high scale and to also suppress the sphalerons to avoid washout while the EW VEV is lowered to its present day value. We first showed the generic ingredients required for a transition to occur at a temperature far above the scale of the zero temperature minimum of a theory. Such findings may be of more general interest.

We then moved onto our specific scenario.
In our example we have demonstrated the combination of: (i) a flavour model, (ii) the symmetry non-restoration effect can give us a novel scenario of high-scale EW baryogenesis. Both the Froggatt-Nielsen  fermions responsible for field dependent Yukawas and the non-restoring scalar degrees-of-freedom combine to give us a strong first order phase transition. The fermions also help to control the symmetry non-restoration effect. Furthermore, the large Yukawas during the phase transition can act as the source of CP violation required to obtain the baryon asymmetry. This naturally allows for an absence of measurable EDMs.

The generic prediction of the scenario is a large number of light scalars, around the EW scale, with a small coupling to the EW Higgs. In our scenario these scalars mix with the SM-like Higgs, although a more complete construction with an extended hidden sector may eventually show that this is not generically necessary. The model presented here may well not be the simplest or most elegant realisation of these ideas, it is presented as a proof-of-principle, we hope it facilitates further exploration of this intriguing possibility.  

\subsubsection*{Acknowledgements}
G. Servant  thanks R. Rattazzi for raising a question after her talk at the 2017 Johns Hopkins workshop that stimulated this work, and is grateful to C. Csaki and D. Kaplan for organising the workshop. We thank N. Suresh for earlier collaboration on phase transitions in the context of FN models and J.R. Espinosa and T. Konstandin for discussions.

\subsubsection*{Note added}
While  this  paper  was  completed, Ref.~\cite{Meade:2018saz} appeared,  which  deals  with similar  ideas,  although  it  focuses  on  the  case  where  the  EW  symmetry  is  not  restored at high temperatures, as it is motivated by GUT/high scale baryogenesis rather than EW baryogenesis.  We also learnt a high scale EW phase transition is being considered by Glioti, Rattazzi, and Vecchi.


\appendix

\section{Coefficients of the mass matrix}
\label{Sec:coefficients}

For completeness, we provide the coefficients of our generalised mass matrix, Eq.~(\ref{eq:fullmassmatrix}). These were found by generating uniformly-distributed pseudorandom numbers with magnitudes in the range $(0.5, \;1.5)$ and phases in the range $(-\pi, \pi)$. We found an initial seed returning approximately the top and charm masses after $\sim \mathcal{O}(1)$ attempts. Some entries were then further adjusted by $\sim 10 \%$ in order to return the mass eigenvalues,
	\begin{equation}
	m_{f1} = 52.9 \; \mathrm{TeV},   \qquad  m_{f2} = 42.4 \; \mathrm{TeV},  \qquad m_{f3}  = 37.3 \; \mathrm{TeV}, 
	\end{equation}        
	\begin{equation}
	m_{f4} = 173 \; \mathrm{GeV}, \qquad  m_{f5} = 1.3 \; \mathrm{GeV},
	\end{equation}
at $v_{\phi}=246$ GeV, $v_{\Delta} = 50$ TeV and $a_{s} = 10$ TeV, where the top and charm correspond to $m_{f4}$ and $m_{f5}$ respectively. The coefficients of the entries proportional to $\phi$ in matrix~(\ref{eq:fullmassmatrix}) are given by
	\begin{equation}
	\begin{pmatrix}
	 0  &  0  & 0  &  0.56 - 1.36i  &	0  	\\
	 0  &  0  & 0  &  0  & 	0 	\\
       	 0  &  0  & 0  &  0  &  0.46 - 0.97i       \\
	 0  &  0  & 0  &  0.40 - 0.51i  &  0 	\\
	 0  &  0  & 0  &  0  &	 0
	\end{pmatrix}.
	\end{equation}
The coefficients of the $\Delta$ entries are given by
		\begin{equation}
	\begin{pmatrix}
	0.71 - 1.13i  &  0  & 0  &  0  &	0  	\\
	 0  &  -0.74 + 0.87i  & 0  &  0  & 	0 	\\
       	 0  &  0  & 0.93 + 0.79i &  0  &  0       \\
	 0.07 + 0.39i  &  0  & 0  &  0  &  0 	\\
	 0  &  0  & 0  &  0  &	 0
	\end{pmatrix}.
	\end{equation}
Finally, the coefficients of the $a_{s}$ entries are given by
	\begin{equation}
	\begin{pmatrix}
	 -1.07 + 1.15i  &  -1.48 + 0.10i  & 0  &  0  &	0  	\\
	 \; \; 0.60 + 0.25i  &  -0.46 + 0.75i  & -0.49 - 0.76i  &  0  & 	0 	\\
       	 0  &  \; \; 0.14 - 0.68i  &  \; \; 0.66 - 0.62i  &  0  &  0       \\
	  \; \; 0.60 + 0.07i  &  \; \; 1.19 - 0.15i  & 0  &  0  &  0 	\\
	 -0.63 + 0.25i  &  0  & 0  &  0  &	 0
	\end{pmatrix}.
	\end{equation}
The mass matrix $\mathcal{M}$, in Eq.~(\ref{eq:fullmassmatrix}), is then formed by summing the three matrices above, multiplied by the relevant field values or $a_{s}$ factor, and dividing by $\sqrt{2}$. We then diagonalised $\mathcal{M}^{\dagger}\mathcal{M}$ at discrete points in field space and then interpolated over these points for reasons of efficiency in the numerical work.

\begin{figure}[t]
\begin{center}
\includegraphics[width=220pt]{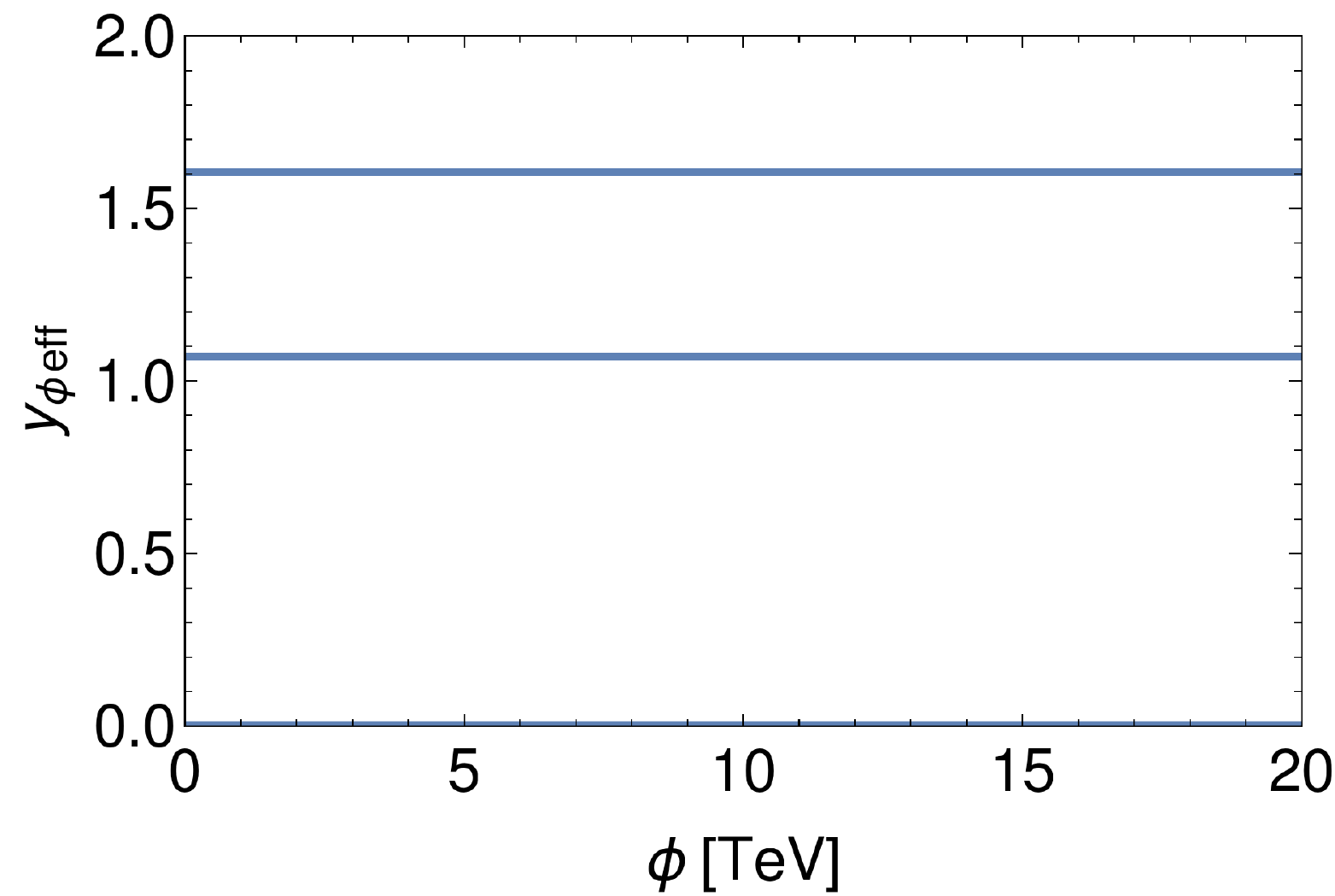} \qquad
\includegraphics[width=220pt]{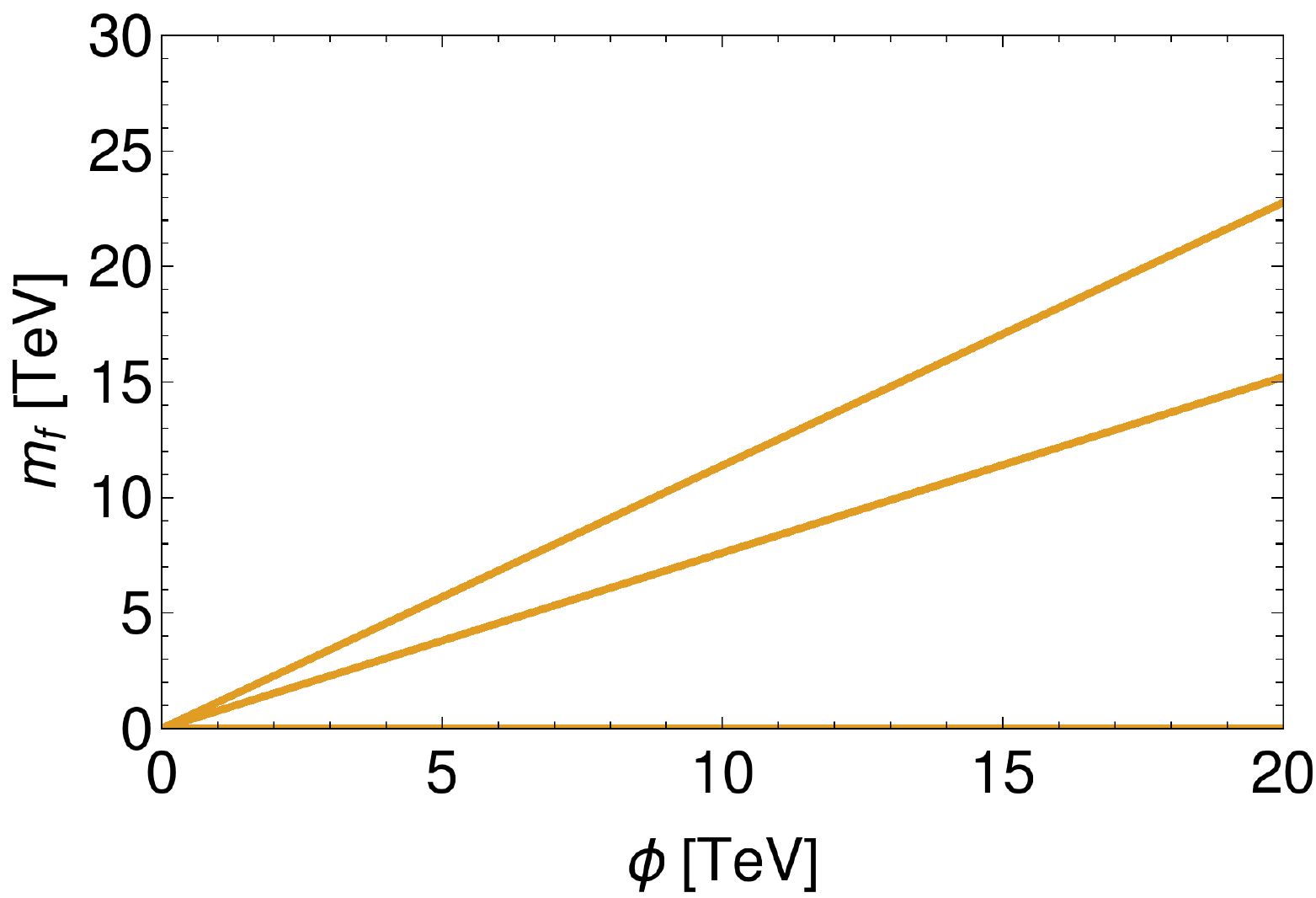}
\end{center}
\caption{\small Left: The effective Yukawa couplings of the electroweak Higgs to the fermions along the $\phi$ axis with no Yukawa variation. Right: The masses of the fermions along the same path.}
\label{fig:yukawasnovar}
\end{figure}

\begin{figure}[t]
\begin{center}
\includegraphics[width=220pt]{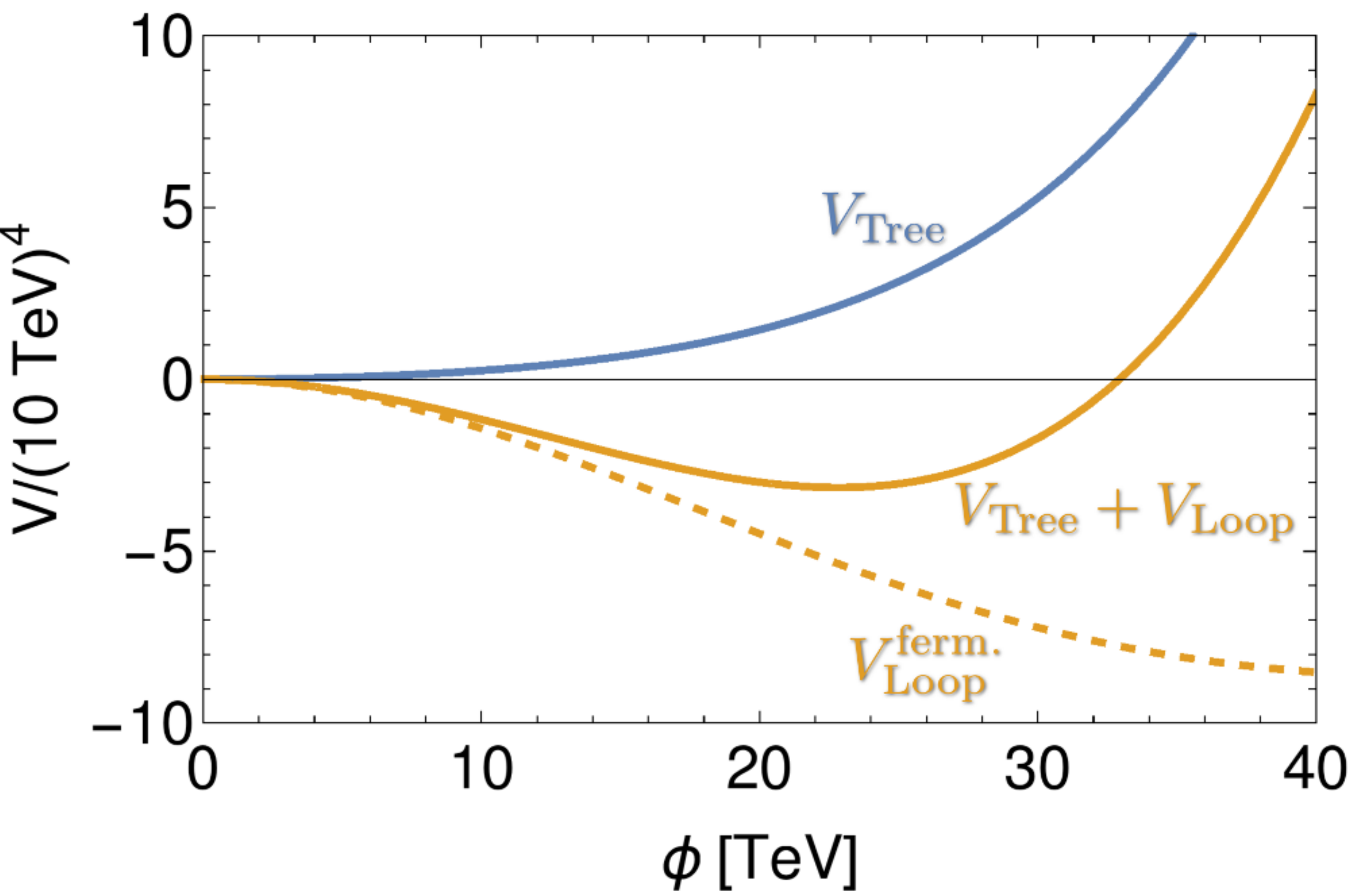}
\includegraphics[width=220pt]{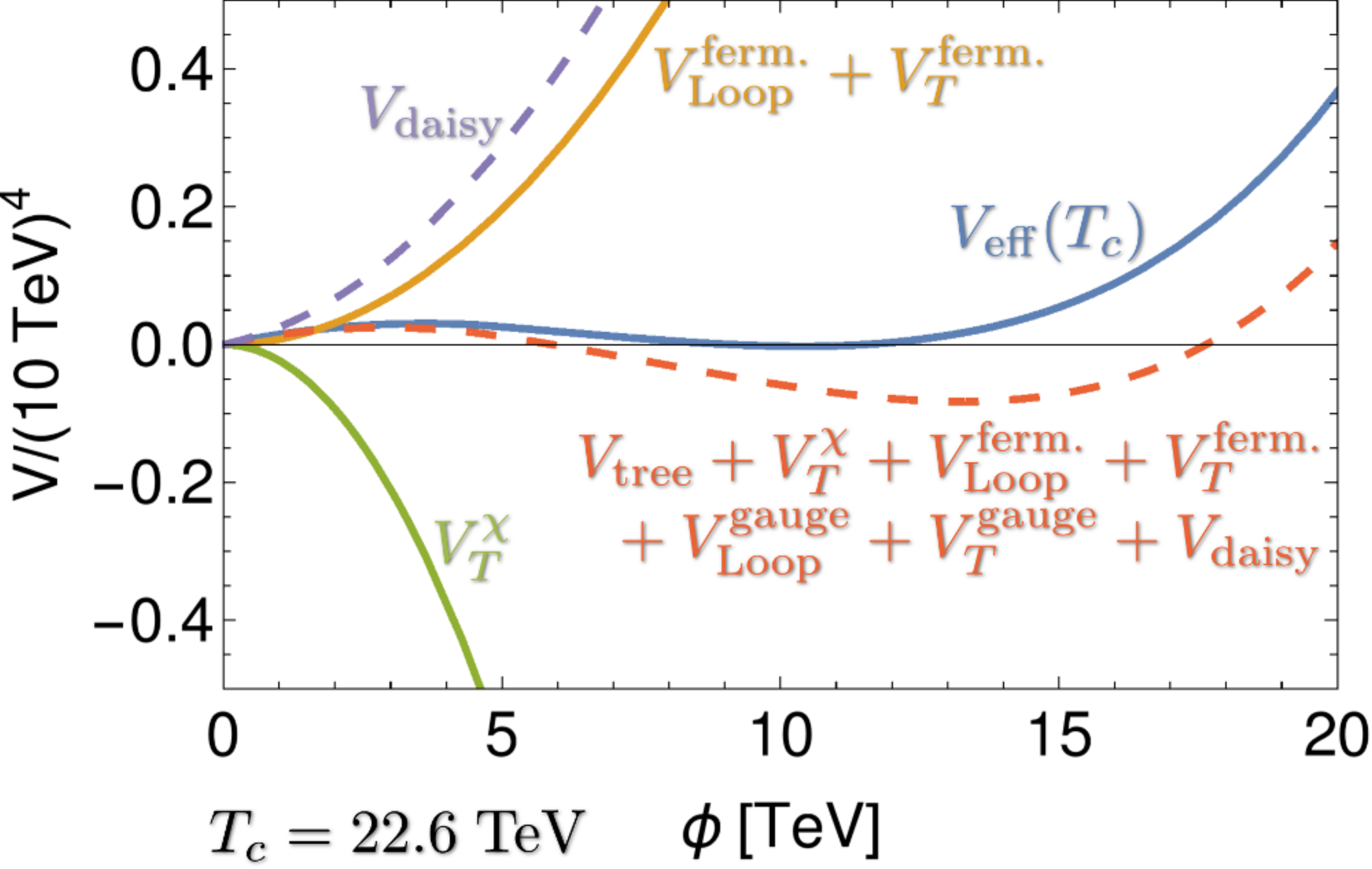}
\end{center}
\caption{\small Left: The one-loop effective potential along the $\phi$ axis at $T=0$ with $a_{s}=0$. Right: The effective potential with $a_{s}=0$ at the EW phase transition critical temperature, now at $T_{c}=22.6$ TeV.}
\label{fig:barriernovar}
\end{figure}

\section{Phase transition with constant Yukawas}
\label{Sec:constyuk}

To contrast with our analysis above, we now consider the Step 1 phase transition with the mixing terms in the fermionic mass matrix switched off, i.e. we set $a_{s}=0$. The Yukawa couplings and masses of the fermions along the, $\Delta=0$, $\phi$ axis are shown in Fig.~\ref{fig:yukawasnovar}. There are three zero mass eigenstates and two with $\mathcal{O}(1)$ couplings to $\phi$. We calculate the critical temperature of the phase transition and find a very weak first order transition at $T_{c}=22.6$~TeV. This is shown in Fig.~\ref{fig:barriernovar}, along with the potential at zero temperature, showing that the qualitative difference to the $a_{s} \neq 0$ case is apparent once the finite $T$ effects are taken into account. For other proposals of using fermions to achieve a strong first order phase transition see~\cite{Baldes:2016rqn,Baldes:2016gaf,Carena:2004ha,Egana-Ugrinovic:2017jib}.

\bibliographystyle{JHEP}   
\bibliography{highscaleewbg}
\end{document}